\documentclass[fleqn,usenatbib]{mnras}

\usepackage{newtxtext,newtxmath}

\usepackage[T1]{fontenc}

\DeclareRobustCommand{\VAN}[3]{#2}
\let\VANthebibliography\thebibliography
\def\thebibliography{\DeclareRobustCommand{\VAN}[3]{##3}\VANthebibliography}

\usepackage{graphicx}	
\usepackage{amsmath}	
\usepackage{multirow}
\usepackage{multicol}

\newcommand{\hb}{H$\beta$}
\newcommand{\hg}{H$\gamma$}

\newcommand{\mgii}{\ion{Mg}{ii}}
\newcommand{\nev}{[\ion{Ne}{v}]}
\newcommand{\oii}{[\ion{O}{ii}]}
\newcommand{\neiii}{[\ion{Ne}{iii}]}
\newcommand{\oiii}{[\ion{O}{iii}]}
\newcommand{\oiiid}{[\ion{O}{iii}]$\lambda\lambda 4959,5007$}
\newcommand{\feii}{\ion{Fe}{ii}}

\newcommand{\sii}{[\ion{S}{ii}]$\lambda\lambda 6717,6731$}
\newcommand{\kms}{\textrm{km}\,\textrm{s}$^{-1}$}

\newcommand{\lsun}{L$_\odot$}
\newcommand{\msun}{M$_\odot$}

\defcitealias{hauschild-roier_differences_2025}{Paper I}

\title[Differences in the NLR of QSOs 1 and 2: II] {The differences in the Narrow Line Region of nearby QSOs 1 and QSOs 2: II - Signatures of kinetic feedback}

\author[Hauschild-Roier et al.]{
Gabriel R. Hauschild-Roier,$^{1}$
Thaisa Storchi-Bergmann,$^{1}$\thanks{E-mail: thaisa.storchi.bergmann@gmail.com}
Rogério Riffel,$^{1}$
Vincenzo Mainieri$^{2}$
and Lara Gatto$^{3}$
\\
$^{1}$Departamento de Astronomia, Universidade Federal do Rio Grande do Sul, Av. Bento Gonçalves 9500, 91501-970, Porto Alegre, RS, Brazil\\
$^{2}$European Southern Observatory, Karl-Schwarzschild-Straße 2, D-
85748 Garching bei Munchen, Germany\\
$^{3}$ Zentrum für Astronomie der Universität Heidelberg, Astronomisches Rechen-Institut, Mönchhofstr 12-14, D-69120 Heidelberg, Germany
}

\date{Accepted XXX. Received YYY; in original form ZZZ}

\pubyear{\the\year{}}

\begin{document}
\label{firstpage}
\pagerange{\pageref{firstpage}--\pageref{lastpage}}
\maketitle

\begin{abstract}
We compare the integrated kinematic properties of the Narrow-Line Region (NLR) of $\sim$1500 type 1 and $\sim$700 type 2 QSOs at $0.4 \le z \le 0.5$ using emission-line profiles from SDSS-IV spectra. The NLR \oiii$\lambda$5007 and H$\beta$ emission lines are fitted using a two-component model comprising a narrow (v$_\mathrm{n}$, $\sigma_\mathrm{n}$) and a broad component (v$_\mathrm{b}$, $\sigma_\mathrm{b}$), while fainter lines are fitted with a single Gaussian component. Both QSO types exhibit similar median narrow-component velocities (v$_\mathrm{n}\approx$\,0) and velocity dispersions ($\sigma_n < 200$\,km\,s$^{-1}$), consistent with motion driven by the host galaxy's gravitational potential. However, parameters associated with the broad, high-velocity gas component  (v$_\mathrm{b}$, v$_{05}$, v$_{10}$, v$_{\nev}$, $\sigma_b$, $W80$ and $W90$) differ: QSOs\,1 display larger blueshifts and broader emission lines than QSOs\,2. This suggests that these components trace Active Galactic Nuclei (AGN) outflows, that appear stronger in QSOs\,1, leading to higher estimated mass-outflow rates ($\dot{\mathrm{M}}_\mathrm{out}$) and kinetic powers ($\dot{\mathrm{E}}_\mathrm{out}$). Rather than offering absolute determinations of $\dot{\mathrm{M}}_\mathrm{out}$ and $\dot{\mathrm{E}}_\mathrm{out}$ -- which remain subject to physical uncertainties -- this statistical study supports the higher NLR excitation found for QSOs\,1 in Paper I, as an orientation effect: in QSOs\,1, we observe gas closer to the AGN core, where velocities and velocity dispersions are intrinsically higher. Nevertheless, the trend of QSOs\,2 exhibiting higher average \oiii\ luminosities than QSOs\,1 suggests an evolutionary contribution, where QSOs\,2 represent an early obscured AGN phase followed by a blowout phase that reveals the central core as a QSO\,1.

\end{abstract}

\begin{keywords}
galaxies: active --  quasars: emission lines -- quasars: general 
\end{keywords}



\section{Introduction} \label{sec:intro}

QSOs are among the most luminous Active Galactic Nuclei (AGN) and are postulated to produce radiative and kinetic feedback on their host galaxies \citep{hopkins_quasar_2010, circosta_super_2018, harrison_agn_2018, trindade_falcao_hubble_2021, villar_martin_interactions_2021, dallagnol_de_oliveira_gauging_2021, audibert_jet-induced_2023, harrison_observational_2024, speranza_multiphase_2024}. Type 1 and type 2 QSOs show characteristics that fit within the Unified Model Scenario \citep{urry_unified_1995}, such as the absence of the Broad Line Region (BLR) and power-law continuum in type 2 sources, while the Narrow Line Region (NLR) is present on both. But is the NLR the same in type 1 and type 2 QSOs?

This is the question we would like to answer, analysing SDSS-IV spectra of a sample of $\sim$1500 QSOs 1 and $\sim$700 QSOs 2 at 0.4$\le$z$\le$0.5. In a previous paper \citep[][hereafter, \citetalias{hauschild-roier_differences_2025}]{hauschild-roier_differences_2025}, we compared the gas excitation of the NLR of the 2 samples and found the following main differences: (1) type 2 QSOs reach, on average, higher \oiii\ luminosities than type 1; (2) type 1 QSOs present higher ionised gas mass -- when traced by H$\beta$ than type 2's; (3) type 1 also present higher \nev/\neiii\ line ratios, showing the presence of higher excitation gas in these objects; (4) type 1 QSOs show stronger signatures of shocks than type 2's. We were also able to calculate the electronic temperature (T$_\mathrm{e}$) for the type 1 QSOs, and found very high values, up to $40\times10^3$\,K. Although the higher T$_\mathrm{e}$ values are expected if you are looking closer to face-on and thus being able to observe NLR gas closer to the ionising source in type 1's as predicted by the Unified Model \citep{antonucci_unified_1993}, the higher \oiii\ luminosity in type 2's is not supported by the Unified Model. Other scenarios are preferred, such as an evolutionary scenario \citep{sanders_ultraluminous_1988, hopkins_black_2005, hopkins_unified_2006} in which a type 2 QSO is initially triggered by interactions and mergers \citep{araujo_nuclear_2023, pierce_galaxy_2023}, giving rise to an obscured nucleus. The radiation field and the AGN winds then proceed to ``wipe clean'' the central region of gas and dust, revealing a type 1 QSO.

However, in \citetalias{hauschild-roier_differences_2025} we looked only at the NLR gas excitation properties. In the present paper we present the comparison between the NLR gas kinematics of QSOs 1 and 2 and discuss how these results fit with the gas excitation.

The NLR gas kinematics is a key tracer of the AGN's kinetic feedback on its host galaxy.  Usage of forbidden emission lines, such as \oiii, often reveals outflows from the ionised gas phase, specially when seen with Integral Field Unit (IFU) spectroscopy data \citep{krajnovic_sauron_2008, storchi-bergmann_bipolar_2018, ruschel-dutra_agnifs_2021, dallagnol_de_oliveira_gauging_2021, tozzi_super_2024}. These outflows can reach large distances from the nucleus, up to $\sim10$\,kpc scales \citep{newman_shocked_2012, newman_sinszc-sinf_2012, dallagnol_de_oliveira_gauging_2021}, and even influence up to $\sim 20$\,kpc, in a kinematically disturbed region (KDR) produced by a combination of the AGN radiation field and winds \citep{gatto_extent_2024}. 

The spatially resolved studies reveal the intrinsic complexity of the kinematics and their physical scales: the combination of a rotation component and additional non-circular components associated, e.g. with inflowing/outflowing gas observed in ionised \citep{krajnovic_sauron_2008, ruschel-dutra_agnifs_2021, dallagnol_de_oliveira_gauging_2021, ilha_active_2022}, atomic, and molecular gas \citep{riffel_mapping_2008, riffel_feeding_2013, hauschild-roier_gas_2022} phases.

The complexity in the calculation of kinematic feedback from AGN, including detailed mapping of the gas kinematics, density and geometry, have been particularly addressed in recent studies by Revalski and collaborators \citep{revalski_quantifying_2021, revalski_quantifying_2022, revalski_quantifying_2025}. \citet{revalski_quantifying_2022}, in particular, highlights the strong radial gradients in velocity, density, and excitation revealed by resolved observations. 

But resolved studies of AGN NLR kinematics as those discussed in the previous two paragraphs, can only be performed in the near Universe. For more distant sources, including the QSOs of our present study, it is necessary to rely on integrated properties of the NLR. The SDSS spectra of our sources were obtained with a 3$^{\prime\prime}$ diameter aperture, corresponding to 16--18\,kpc at the QSOs; we thus only have access to integrated values of the NLR properties. On the other hand, our sample comprises ~2200 sources, and our goal here, rather than perform a detailed mapping of mass outflow rates and powers, is to provide an statistical study in which we compare the {\it integrated} kinematic properties of QSOs 1 to those of QSOs 2. A number of studies of distant AGN have been using single (integrated) spectra to obtain their NLR kinematic properties, including estimates of mass-outflow rates and powers \citep [e.g.][]{mullaney_narrow-line_2013, zakamska_quasar_2014, bischetti_wissh_2017}.

Using integrated spectra of our sources, as well as a similar methodology to that of these previous studies, we aim to look for differences between the kinematic properties of QSOs 1 and 2 and investigate their origin, e.g., if it is consistent with predictions from the Unified Model or point instead to an evolutionary scenario, as done in \citetalias{hauschild-roier_differences_2025}. 


In \citetalias{hauschild-roier_differences_2025} we have already shown how excitation properties derived from the integrated spectra differ between the two samples. In the present paper, we focus on the differences between the {\it kinematic properties}. 


The organization of this paper is as follows: in Section\,2 we review the sample selection and data. In Section\,3, we describe the spectral fitting and analysis, and measurement of kinematic properties. In Sections 4 and 5, we present our results for our Low Luminosity (LL) and High Luminosity (HL) subsamples, respectively. This division in LL and HL subsamples, as done in \citetalias{hauschild-roier_differences_2025}, resulted from the fact that our data comprises two different QSO 2 samples to be compared to one QSO 1 sample. In Section\,6, we compare our findings with those of previous studies and discuss their implications in the context of AGN/QSO feedback and evolution. Finally, in Section\,7, we summarise our results and present our conclusions.

Through this work, we adopted a Flat $\Lambda$CDM cosmology, using H$_0$\,=\,67.4\,\kms\,Mpc$^{-1}$, $\Omega_\Lambda$\,=\,0.685 and $\Omega_m$\,=\,0.315 \citep{planck_collaboration_planck_2020}.

\section{Sample and Data}

For this work, we used the same type 1 and type 2 QSOs samples as for \citetalias{hauschild-roier_differences_2025}, comprised of a total of 1541 QSOs 1 derived from the \citet{lyke_sloan_2020} catalog and 688 QSOs 2, from which 137 originate from \citet{reyes_space_2008} (hereafter Reyes sample) and 551 from \citet{yuan_spectroscopic_2016} (hereafter Yuan sample). The data consist of Sloan Digital Sky Survey (SDSS) spectra from DR16 \citep{ahumada_16th_2020} within the $0.4 \leq z \leq 0.5$ redshift range.

The two QSO 2 samples above present different lower-luminosity L\oiii\ limits: $10^{8.3}$\,L$_\odot$ for the Reyes sample and $10^9$\,L$_\odot$ for the Yuan sample. In order to obtain a proper match between the QSOs 1 and 2 in terms of luminosity limits, the sample was subdivided into the same 2 subsamples utilised in \citetalias{hauschild-roier_differences_2025}:

\begin{enumerate}
    \item a {\it ``low luminosity''} sample (LL), comprised of 1222 QSOs 1 and 73 QSOs 2 originating from the Reyes sample, with \oiii\ luminosities within the range $10^{8.3}$\,L$_\odot$\,<\,L$_{\textrm{\oiii}}<10^{9}$\,L$_\odot$;
    \item a {\it ``high luminosity''} sample (HL), with 319 type 1 objects to be compared with 615 QSO 2 objects, from the combined Reyes and Yuan samples, with L$_{\textrm{\oiii}} > 10^{9}$ L$_\odot$.
\end{enumerate}

The \oiii\ luminosities -- from which we derived also the bolometric luminosity of the AGN -- and corresponding distributions are derived and presented in \citetalias{hauschild-roier_differences_2025}, along with a further description of the sample. We note that the result of the luminosity division was a much larger sample of type 1's than type 2's in the LL sample and the opposite for the HL sample. The peak of the type 1 L\oiii\ distribution is within the LL sample luminosity range, but the peak of the L\oiii\ luminosity distribution seems to be higher for the type 2 than for the type 1 objects. 

\subsection{The effect of the fixed SDSS 3$^{\prime\prime}$ aperture:}

The redshifts of the sources are $0.4<z<0.5$, corresponding to scales 5.4 -- 6.2 kpc arcsec$^{-1}$ at the QSOs. As the diameter of the SDSS fibers is 3$^{\prime\prime}$, the regions covered at the galaxies have extents of 16.3 -- 18.5 kpc, corresponding to radii in the range 8.2 -- 9.3\,kpc, thus most probably encompassing the whole NLR or at least most of it in our sources. The NLR is thus not resolved in our measurements and our results along the paper correspond to $integrated$ properties.

\section{Methodology}

\subsection{Spectral fits}\label{sec:specfit}

\subsubsection{Emission line fitting}

Following the methodology presented in \citetalias{hauschild-roier_differences_2025}, the spectra were fitted using the Python library PyQSOFit \citep{guo_pyqsofit_2018}. In the fit of the NLR emission lines, we assigned two kinematic components to the strongest lines (\oiiid, \hb, \hg\ and \mgii): a narrow, NLR$_\mathrm{n}$, and a broad component, NLR$_\mathrm{b}$.  As adopted in previous studies \citep[e.g.][]{heckman_emission-line_1981, greene_comparison_2005, mullaney_narrow-line_2013, zakamska_quasar_2014}, the narrow component is assumed to probe gas orbiting in the gravitational field of the galaxy (e.g., circular motions), while the broad component probes the non-circular motions, most probably outflows from the AGN. The weakest emission lines (\oiii$\lambda4363$, \neiii, \oii$\lambda3728$ and \nev) were fitted with only the narrow NLR component. The permitted lines of the BLR -- \hb, \hg\ and \mgii\ -- were fitted with two components. 

Each component was assumed to have a Gaussian profile. In order to simplify the computation, we constrained each kinematic component within 4 fitting groups, defined according to the proximity in wavelengths of the lines and/or same physical constraints. Group (1) comprises only the \mgii$\lambda$2800 profile; group (2) comprises only the \nev\ line; group (3) the \oii$\lambda3728$ and \neiii\ lines; group (4) the \hg, \oiii$\lambda4363$, \hb\ and \oiiid\ emission lines. A full description of this procedure is found in Section\,3 of \citetalias{hauschild-roier_differences_2025}.

\subsubsection{Kinematic properties of the NLR}

As mentioned above, the emission line fitting of the strongest NLR emission lines, \oiii\ and H$\beta$, was performed using two Gaussian components -- a narrow, characterized by peak velocity v$_\mathrm{n}$ and velocity dispersion $\sigma_n$ and a broad, characterized by v$_\mathrm{b}$ and $\sigma_b$.

Besides the above, we have also obtained other kinematic properties: the ``non-parametric'' quantities v$_{05}$, v$_{10}$, v$_{90}$, v$_{95}$, $W80$ (v$_{90}$ -- v$_{10}$) and $W90$ (v$_{95}$ -- v$_{05}$) \citep{zakamska_quasar_2014}. The first four parameters were used as tracers of negative and positive asymmetry of the line profiles, while $W80$ and $W90$ are indicators of the velocity dispersion at the highest velocities and usually trace non-circular motions such as outflows or inflows. The first four quantities are obtained by the integration of the flux density in velocity space: they are the upper velocity limits of the integral resulting in 5\%, 10\%, 90\% and 95\% of the total integrated flux, respectively. The ``zero'' velocity is defined by the rest wavelength of the emission line. 
Finally, as a tracer of the median velocity of the emission line profile, we also compute the centroid velocity $\langle v \rangle$, defined as the expected value of the velocity weighed by the flux density.

\begin{figure}
    \centering
    \includegraphics[width=0.96\columnwidth]{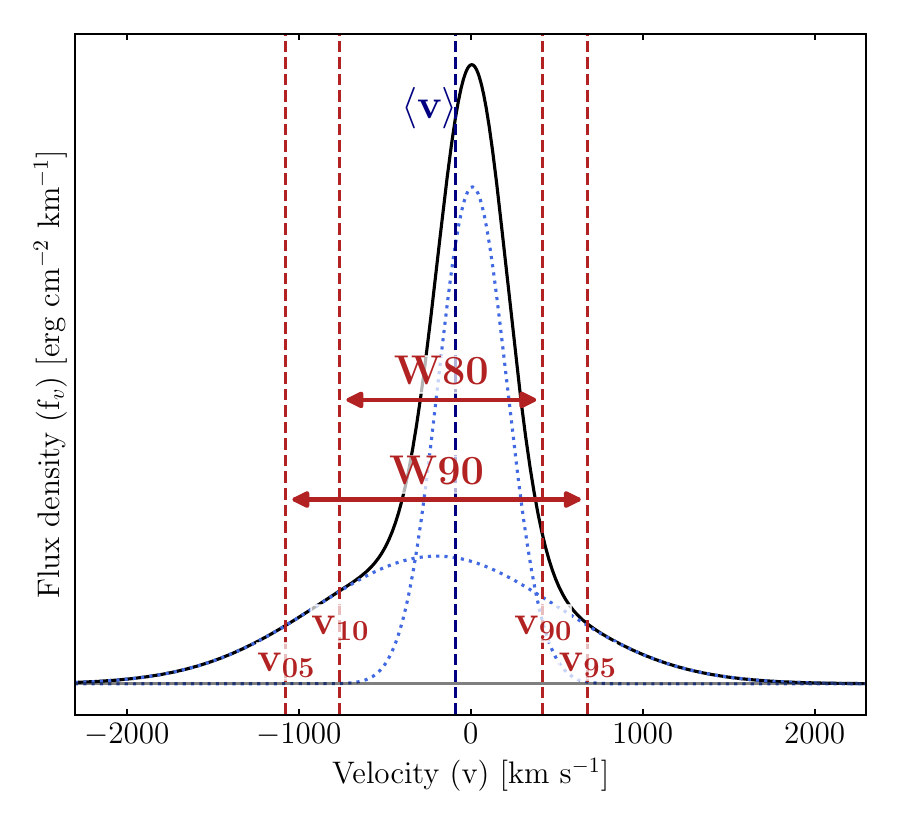}
    \caption{Schematic view of $v_{05}$, $v_{10}$, $v_{90}$, $v_{95}$, $W80$, $W90$ and $\langle v \rangle$ overplotted on a double Gaussian emission line fit. The single Gaussian components are shown in dotted light blue curves, and the full profile in black. As vertical red dashed lines, we denote the $v_{05}$, $v_{10}$, $v_{90}$ and $v_{95}$ values, with a red arrow between them showing the profile widths $W80$ and $W90$ for this example profile. Finally, as a dark blue dashed vertical line we show the centroid velocity $\langle v \rangle$.}
    \label{fig:parameters}
\end{figure}

We show a schematic view of these quantities in Figure\,\ref{fig:parameters} plotted over an example of a two component Gaussian fit. In this figure, the individual Gaussian lines are shown as light blue dotted curves and the full emission line profile as a continuous black curve. The $v_{05}$, $v_{10}$, $v_{90}$ and $v_{95}$ values for the profile are indicated by vertical dashed red lines. The differences between these values, $W80$ and $W90$, are denoted as red arrows. Finally, the centroid velocity $\langle v \rangle$ is marked as a dark blue dashed vertical line.

Since our aim is to look for similarities and differences between the NLRs of the two QSO populations, for each NLR parameter obtained, we perform a Kolmogorov-Smirnov (KS) test in the distributions of the values of these parameters for the QSOs 1 and 2, under the assumption that the two distributions are drawn from the same parent distribution. We adopt $\alpha=0.05$ as the significance level of the test throughout this work (above which the distributions are considered similar), as we have done in \citetalias{hauschild-roier_differences_2025}.

\subsection{Estimated mass outflow rates and powers}
\label{sec:methods_outflow}

Besides comparing the integrated kinematic properties described in the previous Section for QSOs 1 and 2, we have also estimated mass-outflow rates $\dot{\mathrm{M}}_\mathrm{out}$ and powers $\dot{\mathrm{E}}_\mathrm{out}$ as tracers of the rate of expulsion of the gas from the central regions and of the kinetic power injected into the gas by the AGN, respectively. Rather than precise estimates of this quantities, our goal is to compare these ``kinematic feedback indicators" with those obtained in previous studies using similar methodology \citep[e.g.][]{fiore_agn_2017, gatto_extent_2024, dallagnol_de_oliveira_gauging_2021}. In addition, we aim to compare the kinematic properties to the excitation properties obtained in \citetalias{hauschild-roier_differences_2025}.

In the choice of relevant parameters, we use as main reference for our estimates the paper of \citet{dallagnol_de_oliveira_gauging_2021}, in which a resolved study is performed for 8 QSOs 2 from the same \citet{reyes_space_2008} sample we have used, having similar luminosities, although most of them have a somewhat lower redshift than ours (0.2--0.4). The assumptions we had to make along the paper, due to the lack of resolution of our data, are based on this resolved study.


As a means to investigate the effect of different methodologies and assumptions on the calculation of 
$\dot{\mathrm{M}}_\mathrm{out}$ and $\dot{\mathrm{E}}_\mathrm{out}$, we choose to use three  methods \citep[e.g.][]{liu_observations_2013, storchi-bergmann_bipolar_2018, dallagnol_de_oliveira_gauging_2021, gatto_extent_2024}. 

These methods are described below: (1) we first assume that the outflow velocity is traced by the broad NLR component; (2) we then use $W80$ and (3) $W90$  as the outflow velocities.

\subsection*{Broad component method}

This first method is based on the assumption that the outflows are traced by the broad component of the NLR in the \oiii\ and \hb\ emission-line profiles. The velocity of the outflow is assumed to be close to the maximum velocity value in the broad profile \citep[e.g.][]{rupke_multiphase_2013,fiore_agn_2017,dallagnol_de_oliveira_gauging_2021} 
being defined as:

\begin{equation}
    \mathrm{v_{out,broad}} = |\mathrm{v_b}| + 2\sigma_\mathrm{b}
\end{equation}

\noindent in which $\mathrm{v_b}$ and $\sigma_\mathrm{b}$ are the centroid velocity of the broad component relative to the systemic velocity of the galaxy and velocity dispersion of the broad component of the \hb\ line, respectively. From this definition, we then calculate the mass outflow rate $\dot{\mathrm{M}}_\mathrm{out, broad}$ using the following expression:

\begin{equation}
    \dot{\mathrm{M}}_\mathrm{out, broad} = \frac{\mathrm{M_b v_{out,broad}}}{R}
\end{equation}

\noindent where $\mathrm{M_b}$ is the mass corresponding to the emission in the broad NLR component of \hb\ -- derived via equation 1 in  \citetalias{hauschild-roier_differences_2025}, for an assumed geometry of a filled bicone with half-opening angle of 45$^\circ$
\citep{shimizu_multiphase_2019, dallagnol_de_oliveira_gauging_2021},
and $R$ is the radius of the outflow. 

\subsubsection{Correction for reddening}

In \citetalias{hauschild-roier_differences_2025} we opted not to correct our measurements for reddening due to the selected redshift range that did no cover the H$\alpha$ line in the SDSS spectra, and the H$\gamma$ line had low signal-to-noise ratio in most sources. But here, as we aim to estimate mass-outflow rates and powers and compare our values with those from other authors, we opted to make a correction based on the reddening measurements from \citet{dallagnol_de_oliveira_gauging_2021}, which support previous measurements from \citet{reyes_space_2008} for QSOs 2 and \citet{lu_reddening_2019} for QSOs 1. These three papers concur on a mean NLR reddening of $E(B-V)\approx 0.4$ and with a mean variation of $\approx$0.2. We have thus corrected all luminosities for $E(B-V)=0.4$, and considered the mean variation of $\approx$0.2 in the uncertainties discussed in Sec.\,\ref{uncertainties}.

\subsubsection{The extent of the outflow}

There is no spatial information in our spectroscopic data, as the SDSS fiber aperture radius is equivalent to 8.2 -- 9.3\,kpc at the studied redshift range. We have thus used, as reference (as already pointed out above), the resolved study of \citet{dallagnol_de_oliveira_gauging_2021}, to estimate the extent of the outflow $R$. These authors found, similarly to \citet{fischer_hubble_2018}, that the outflow extent is a fraction of the extent of the NLR, with a mean value of $\approx$ 50\% for (our) higher luminosity sample \citep{dallagnol_de_oliveira_gauging_2021}.

The extent of NLR R$_{NLR}$, in turn, has been shown, for the same QSO 2 sample, by \citet{storchi-bergmann_bipolar_2018} to be proportional to the \oiii\ luminosity, according to the equation:

\begin{equation}
\log(R_{\text{NLR}}) = -18.12 + \log L_{\text{[O III]}}
\end{equation}

A similar relation was obtained by \citet{fischer_hubble_2018}, although for a lower luminosity sample. The extent $R$ of the outflow will then be $R=0.5 R_{NLR}$.


Finally, we can derive the mass outflow power $\dot{\mathrm{E}}_\mathrm{out}$ utilising the expression from \citet{dallagnol_de_oliveira_gauging_2021}:

\begin{equation}
\label{eq:edot}
    \dot{\mathrm{E}}_\mathrm{out,broad} = \frac{1}{2} \dot{\mathrm{M}}_\mathrm{out,broad} \mathrm{v_{out,broad}^2}
\end{equation}

\subsection*{$W80$ and $W90$ methods}

The second method to obtain $\dot{M}$ and $\dot{E}$ assumes that the outflow is traced by $W80$, so that it corresponds to the group velocity of the outflowing gas. Some of the advantages of $W80$ are that it is not dependent of any models utilised for data fitting and that it is not sensitive to the SNR of the broad component, but to that of the whole \hb\ profile \citep[e.g.][]{liu_observations_2013, sun_sizes_2017}. We utilize the following relation to derive the mass outflow rate in this case:

\begin{equation}
    \dot{\mathrm{M}}_\mathrm{out,W80} = \frac{\mathrm{M\ W80}}{R}
\end{equation}

\noindent where we included the total \hb\ mass in the calculation, also derived from the expression presented in \citetalias{hauschild-roier_differences_2025}.  This relation is similar to the one utilised by \citet{ruschel-dutra_agnifs_2021}, differing only in the use of $W80$ as the outflow velocity, instead of $W80/2$. From this, we derive the mass outflow power $\dot{\mathrm{E}}_\mathrm{out,W80}$ utilising an expression derived from Equation\,\ref{eq:edot}:

\begin{equation}
    \dot{\mathrm{E}}_\mathrm{out,W80} = \frac{1}{2} \dot{\mathrm{M}}_\mathrm{out,W80} \mathrm{W80}^2
\end{equation}


The third method is identical to the second, except for the use of $W90$ as the outflow tracer instead of $W80$ since it comprises a larger portion of the emission line profile, reaching more extreme velocities \citep{zakamska_quasar_2014}. The expressions are the same as the above, only replacing $W80$ by $W90$.




\subsection*{Coupling efficiency}

In addition to $\dot{\mathrm{M}}_\mathrm{out}$ and $\dot{\mathrm{E}}_\mathrm{out}$, we also estimate the bolometric luminosity L$_\mathrm{bol}$ with the following relation from \citet{trump_biases_2015}:

\begin{equation}
    \frac{\mathrm{L}_\mathrm{bol}}{10^{40}\ \mathrm{erg\ s^{-1}}} = 112 \left(\frac{\mathrm{L}(\oiii)}{10^{40}\ \mathrm{erg\ s^{-1}}} \right)^{1.2}
\end{equation}

\noindent where $\mathrm{L}(\oiii)$ is the luminosity of the \oiii$\lambda$5007 emission line. 

Using it, we define the feedback coupling efficiency $\varepsilon$ as the ratio between the kinetic power and bolometric luminosity \citep[e.g.][]{di_matteo_energy_2005, hopkins_quasar_2010}:

\begin{equation}
    \varepsilon = \frac{\dot{\mathrm{E}}_\mathrm{out}}{\mathrm{L}_\mathrm{bol}}
\end{equation}

This parameter is used to gauge the fraction of the AGN luminosity that is transformed into kinetic power, which will produce ``feedback" on the host galaxy, via heating and/or expulsion of gas that may halt or delay star formation in the galaxy, at least in the nuclear region.
This coupling efficiency can then be compared to minimum threshold values predicted by models for producing effective feedback on the galaxy. For example, \citet{di_matteo_energy_2005} proposed a threshold value of $\varepsilon = 0.04$, while \citet{hopkins_quasar_2010} proposed a smaller value $\varepsilon = 0.005$. Other simulations, like IllustrisTNG propose higher values, of $\varepsilon = 0.1$ \citep{weinberger_simulating_2017}.

\section{Results - Low Luminosity Sample (LL)}

In this section, we present the results regarding the LL subsample, comparing the kinematic properties of QSOs 1 and 2. First, we discuss the NLR kinematic parameters obtained from the line-profile fits, and then the derived outflow properties $\dot{\mathrm{M}}_\mathrm{out}$ and $\dot{\mathrm{E}}_\mathrm{out}$.

\subsection{NLR kinematic parameters from the line-profile fits} \label{sec:kin_res_lowL}

\begin{figure*}
    \centering
    \includegraphics[width=0.8\textwidth]{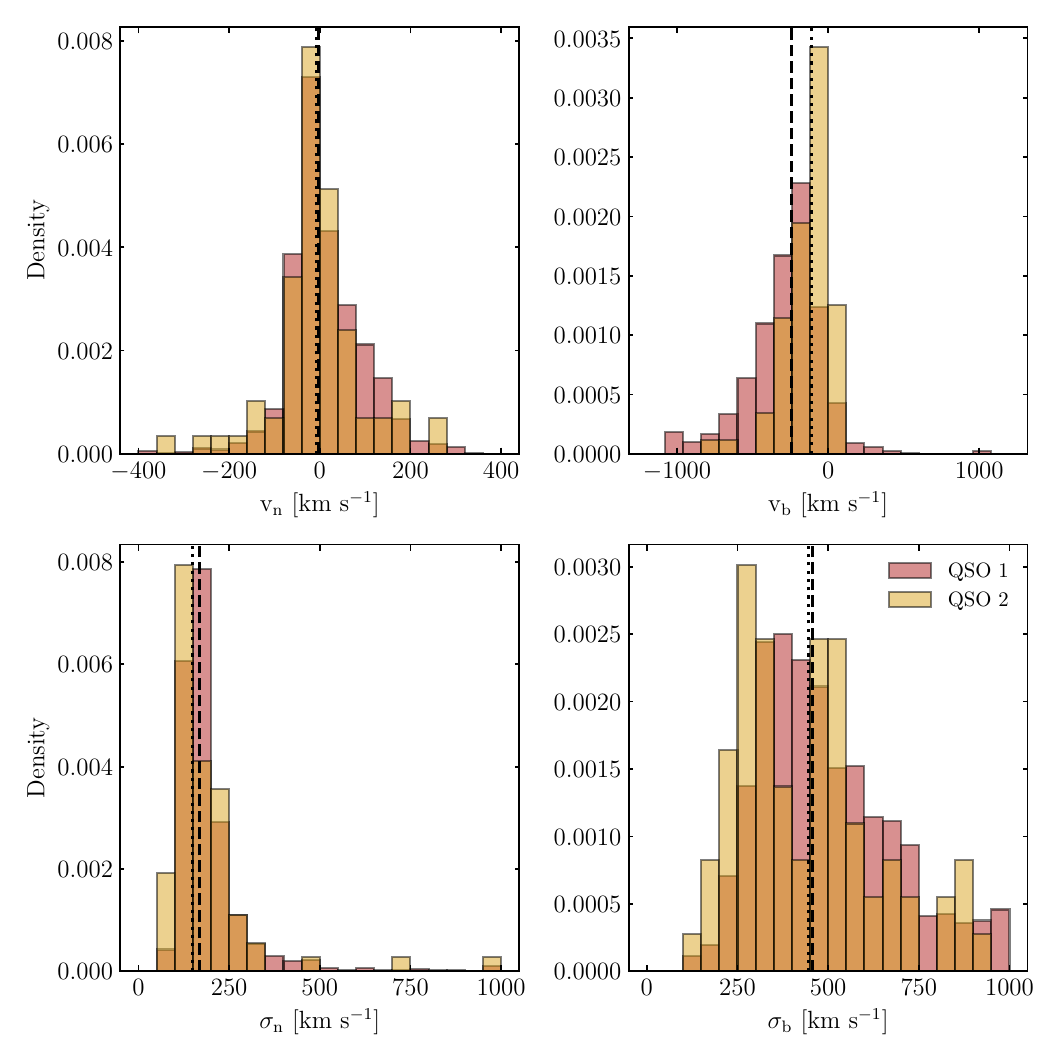}
    \caption{Normalized distributions of the velocity components v$_n$ and v$_b$ (top panels) and corresponding velocity dispersions $\sigma_n$ and $\sigma_b$ (bottom panels) for the \oiii$\lambda$5007 emission line, for the LL sample.  The QSO 1 and QSO 2 populations are shown as red and yellow histograms, respectively, with the median of each population shown as black dashed (QSO 1) or dotted (QSO 2) vertical black lines.}
    \label{fig:hist_voiii_comp_lowL}
\end{figure*}

\begin{figure*}
    \centering
    \includegraphics[width=\textwidth]{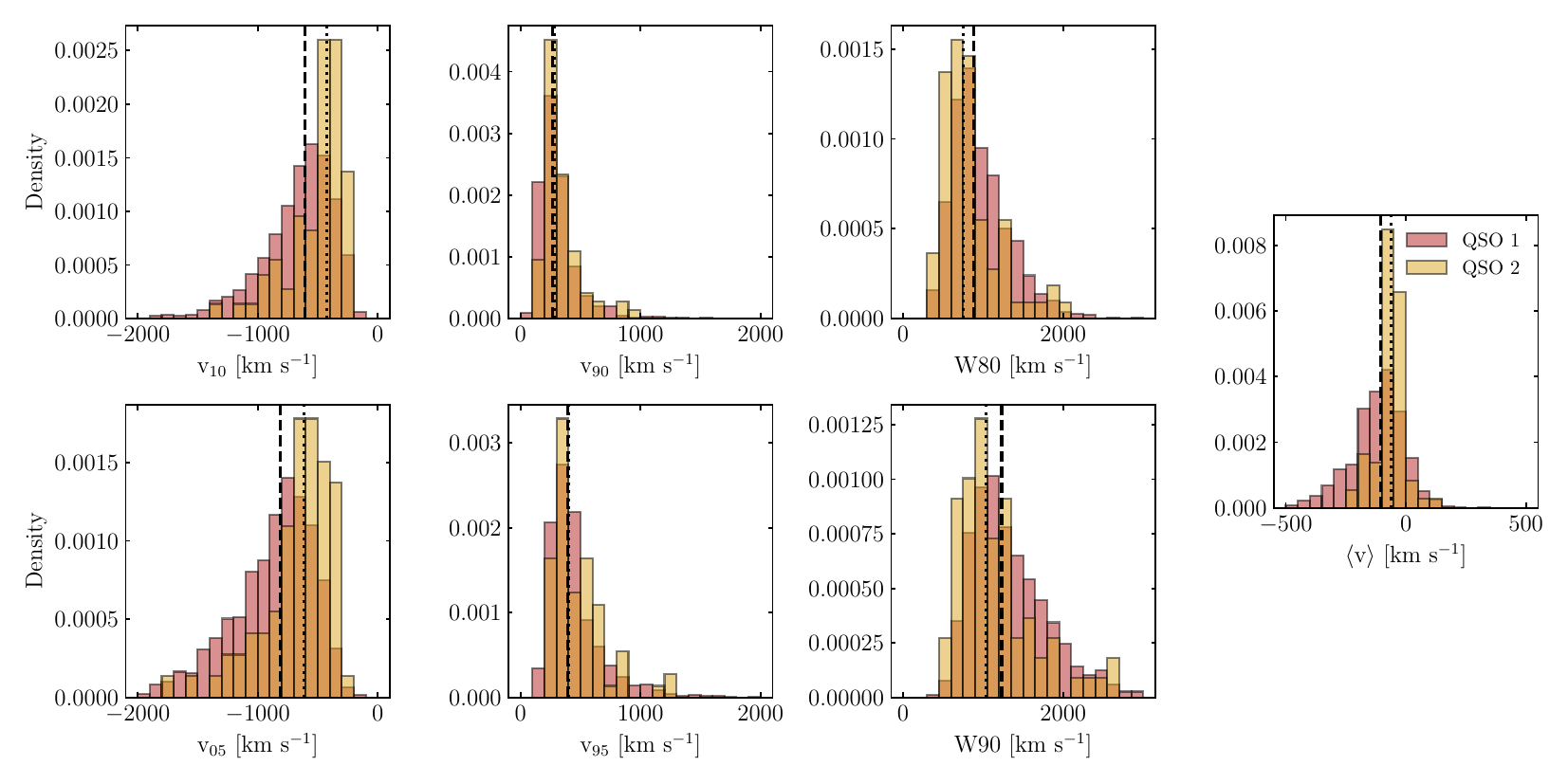}
    \caption{Normalized v$_{10}$, v$_{90}$, W80 (top row), v$_{05}$, v$_{95}$, W90 (bottom row) and $\langle$v$\rangle$ (right panel) distributions of the \oiii$\lambda$5007 emission line of the LL sample, from top to bottom. The QSO 1 and QSO 2 populations are shown as red and yellow histograms, respectively, showing the median of each population as black dashed or dotted vertical black lines.}
    \label{fig:hist_w80_lowL}
\end{figure*}

\begin{figure*}
    \centering
    \includegraphics[width=0.8\textwidth]{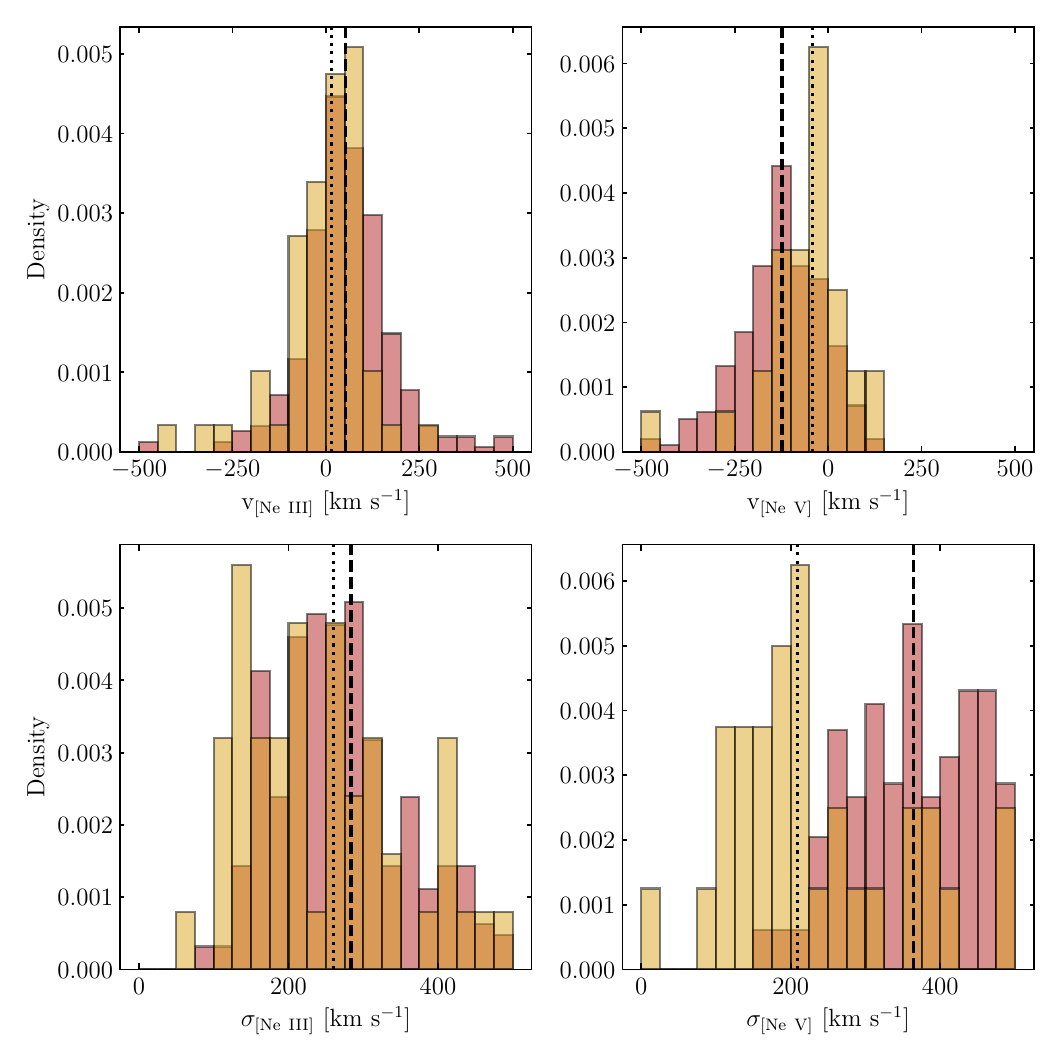}
    \caption{Normalized distributions of the velocity v (top) and velocity dispersion $\sigma$ (bottom) from the \neiii\ (left) and \nev\ (right) emission lines  of the LL sample. The QSO 1 and QSO 2 populations are shown as red and yellow histograms, respectively, showing the median of each population as black dashed and dotted vertical black lines, respectively.}
    \label{fig:hist_vnev_comp_lowL}
\end{figure*}

In this section we discuss the values of the parameters derived from the NLR emission-line profile fits: v$_\mathrm{n}$, $\sigma_n$, v$_ b$, $\sigma_b$, $W80$ and $W90$, comparing their distributions for the QSOs 1 and 2. The results are shown in Figs.\,\ref{fig:hist_voiii_comp_lowL} -- \ref{fig:mdot_edot_lowL} and the median values of the parameters and measurements are listed in Table \ref{tab:median_kin_lowL}.

\begin{table}
\centering
\caption{Median values for the kinematic parameters, mass outflow rate, outflow power and bolometric luminosities of QSOs 1 and 2 for the LL sample. Units are: \kms for velocities and dispersions, \msun\,yr$^{-1}$ for $\dot{\mathrm{M}}_\mathrm{out}$, erg\,s$^{-1}$ for $\dot{\mathrm{E}}_\mathrm{out}$ and L$_\mathrm{bol}$, and kpc for R$_\mathrm{out}$. We also include the KS test p-value for the difference between the QSOs 1 and 2 distributions for each parameter.}
\label{tab:median_kin_lowL}
\begin{tabular}{l|ccc}
\hline \hline
Parameter & QSO 1 & QSO 2 & p$_\mathrm{KS}$ \\
\hline
v \oiii$\lambda$5007,n & -3.16 & -7.76 & 2.80e-01 \\
$\sigma$ \oiii$\lambda$5007,n & 168.94 & 150.24 & 9.90e-04 \\
v \oiii$\lambda$5007,b & -241.70 & -107.81 & 1.08e-09 \\
$\sigma$ \oiii$\lambda$5007,b & 456.06 & 445.47 & 8.17e-03 \\
v \hb,n & -6.50 & -22.74 & 7.66e-02 \\
$\sigma$ \hb,n & 181.49 & 151.79 & 3.38e-02 \\
v \hb,b & -263.94 & -77.19 & 2.34e-09 \\
$\sigma$ \hb,b & 482.38 & 375.82 & 2.25e-03 \\
v \oiii$\lambda$4363 & -31.68 & -19.10 & 4.42e-01 \\
$\sigma$ \oiii$\lambda$4363 & 354.50 & 219.19 & 1.76e-01 \\
v \neiii$\lambda$3869 & 51.35 & 14.10 & 1.82e-03 \\
$\sigma$ \neiii$\lambda$3869 & 283.53 & 259.98 & 2.46e-02 \\
v \oii$\lambda$3728 & 6.84 & -16.31 & 2.89e-05 \\
$\sigma$ \oii$\lambda$3728 & 233.92 & 223.45 & 1.79e-05 \\
v \nev$\lambda$3426 & -172.61 & -44.74 & 1.94e-05 \\
$\sigma$ \nev$\lambda$3426 & 363.72 & 214.03 & 1.67e-18 \\
v$_{05}$ & -810.50 & -612.37 & 3.60e-07 \\
v$_{10}$ & -604.88 & -427.40 & 2.52e-08 \\
v$_{90}$ & 268.77 & 283.74 & 7.08e-02 \\
v$_{95}$ & 393.28 & 400.46 & 9.18e-02 \\
W80 & 885.93 & 754.83 & 2.83e-04 \\
W90 & 1231.62 & 1035.58 & 4.11e-03 \\
$\langle$v$\rangle$ & -106.55 & -63.02 & 3.00e-09 \\
$\dot{\mathrm{M}}_\mathrm{out, broad}$ & 9.58 & 4.95 & 1.82e-06 \\
$\log(\dot{\mathrm{E}}_\mathrm{out, broad})$ & 42.61 & 42.20 & 1.52e-06 \\
$\dot{\mathrm{M}}_\mathrm{out, W80}$ & 8.08 & 4.23 & 3.63e-07 \\
$\log(\dot{\mathrm{E}}_\mathrm{out, W80})$ & 42.25 & 41.87 & 8.17e-06 \\
$\dot{\mathrm{M}}_\mathrm{out, W90}$ & 11.29 & 6.09 & 1.43e-07 \\
$\log(\dot{\mathrm{E}}_\mathrm{out, W90})$ & 42.70 & 42.37 & 3.06e-05 \\
log(L$_\mathrm{bol})$ & 45.22 & 45.36 & 6.72e-03 \\
R$_\mathrm{out}$ & 1.28 & 1.40 & 6.72e-03 \\
\hline
\end{tabular}
\end{table}

\begin{table}
\centering
\caption{Median values for the kinematic parameters, mass outflow rate, outflow power and bolometric luminosities of type 1 and type 2 QSOs for the HL sample. Units are: \kms for velocities and dispersions, \msun\,yr$^{-1}$ for $\dot{\mathrm{M}}_\mathrm{out}$, erg\,s$^{-1}$ for $\dot{\mathrm{E}}_\mathrm{out}$ and L$_\mathrm{bol}$, and kpc for R$_\mathrm{out}$. We also include the KS test p-value for the difference between the QSOs 1 and 2 distributions for each parameter.}
\label{tab:median_kin_highL}
\begin{tabular}{l|ccc}
\hline \hline
Parameter & QSO 1 & QSO 2 & p$_\mathrm{KS}$ \\
\hline
v \oiii$\lambda$5007,n & -23.29 & -4.15 & 3.69e-09 \\
$\sigma$ \oiii$\lambda$5007,n & 186.38 & 181.29 & 2.81e-01 \\
v \oiii$\lambda$5007,b & -161.77 & -60.34 & 3.33e-19 \\
$\sigma$ \oiii$\lambda$5007,b & 595.07 & 443.49 & 9.88e-22 \\
v \hb,n & -28.02 & -3.87 & 1.84e-11 \\
$\sigma$ \hb,n & 185.05 & 182.46 & 5.56e-01 \\
v \hb,b & -199.99 & -55.66 & 1.94e-20 \\
$\sigma$ \hb,b & 568.54 & 438.00 & 4.54e-12 \\
v \oiii$\lambda$4363 & -32.41 & -7.53 & 4.33e-06 \\
$\sigma$ \oiii$\lambda$4363 & 201.10 & 191.17 & 3.71e-02 \\
v \neiii$\lambda$3869 & 64.02 & 68.17 & 3.28e-04 \\
$\sigma$ \neiii$\lambda$3869 & 266.09 & 221.47 & 3.58e-12 \\
v \oii$\lambda$3728 & -8.79 & -6.45 & 4.53e-08 \\
$\sigma$ \oii$\lambda$3728 & 277.50 & 258.63 & 2.03e-09 \\
v \nev$\lambda$3426 & -52.83 & -10.22 & 1.39e-27 \\
$\sigma$ \nev$\lambda$3426 & 301.60 & 234.84 & 1.53e-92 \\
v$_{05}$ & -744.06 & -521.38 & 9.55e-19 \\
v$_{10}$ & -505.22 & -380.71 & 8.78e-18 \\
v$_{90}$ & 320.85 & 316.06 & 1.21e-01 \\
v$_{95}$ & 492.05 & 436.38 & 1.83e-04 \\
W80 & 865.57 & 693.18 & 1.35e-11 \\
W90 & 1276.81 & 960.15 & 4.91e-15 \\
$\langle$v$\rangle$ & -70.37 & -23.07 & 2.89e-26 \\
$\dot{\mathrm{M}}_\mathrm{out, broad}$ & 11.49 & 6.63 & 2.49e-19 \\
$\log(\dot{\mathrm{E}}_\mathrm{out, broad})$ & 42.83 & 42.29 & 5.70e-29 \\
$\dot{\mathrm{M}}_\mathrm{out, W80}$ & 11.08 & 6.97 & 5.48e-24 \\
$\log(\dot{\mathrm{E}}_\mathrm{out, W80})$ & 42.37 & 42.04 & 1.49e-14 \\
$\dot{\mathrm{M}}_\mathrm{out, W90}$ & 15.56 & 9.73 & 1.59e-25 \\
$\log(\dot{\mathrm{E}}_\mathrm{out, W90})$ & 42.90 & 42.46 & 2.97e-18 \\
log(L$_\mathrm{bol})$ & 45.94 & 45.92 & 6.37e-02 \\
R$_\mathrm{out}$ & 2.44 & 2.42 & 6.37e-02 \\
\hline
\end{tabular}
\end{table}

\vspace{0.2cm}

\noindent{\bf\oiii$\lambda5007$ -- } The panels of Figure\,\ref{fig:hist_voiii_comp_lowL} show the distribution of velocities  v (top) and velocity dispersion $\sigma$ (bottom) values for the narrow (v$_\mathrm{n}$, $\sigma_n$) and broad (v$_\mathrm{b}$, $\sigma_b$) components of the \oiii$\lambda$5007 line, comparing the distributions for the QSOs 1 (red) and QSOs 2 (yellow).
The corresponding median values for each distribution are shown in Table\,\ref{tab:median_kin_lowL}. 

The velocity distributions are similar for v$_\mathrm{n}$, with both types centred at zero but with a larger asymmetry towards redshifted velocities. The opposite is observed for the broad component v$_\mathrm{b}$, whose distributions are skewed towards blueshifted values and centred at negative velocities for both populations, with a larger blueshift observed in QSOs 1.

The velocity dispersions distributions, for both the narrow and broad components and type 1 and 2 QSOs, are asymmetric with respect to the median, presenting a tail towards the highest values. The $\sigma_\mathrm{n}$ distribution looks similar for both types, typically not reaching values larger than 500\,\kms, while the $\sigma_\mathrm{b}$ distribution reaches values up to 1000\,\kms\ for both populations. 

For every panel, a KS-test was performed in order to test the similarity between the QSOs 1 and 2 populations, with the corresponding p$_{KS}$ values presented in the rightmost column of Table\,\ref{tab:median_kin_lowL}. In spite of the median values for the distributions being very similar, for all parameters it shows that there is statistical difference between the populations, except for v$_\mathrm{n}$, whose p$_\mathrm{KS} = 0.28$, larger than the adopted significance level of 0.05.

We present the v$_{10}$, v$_{90}$ and W80 distributions for the \oiii$\lambda$5007 profile in the top panels of Figure\,\ref{fig:hist_w80_lowL}, along with v$_{05}$, v$_{95}$ and W90 in the bottom panels. The $\langle$v$\rangle$ distributions are shown in the rightmost panel.


The leftmost panels reveal the prevalence of 
blueshift-skewed emission-line profiles -- as traced by both v$_{05}$ and v$_{10}$ when compared to v$_{90}$ and v$_{95}$ -- for both the QSOs 1 and 2: in other words, the negative velocities (v$_5$ and v$_{10}$) reach higher absolute values than the positive velocities (v$_{90}$ and v$_{95}$) for both QSOs 1 and 2. The type 1s show more blueshifted median v$_{05}$ and v$_{10}$ values when compared to type 2s, as listed in Table\,\ref{tab:median_kin_lowL}. The median values of the v$_{90}$ and v$_{95}$ distributions of QSOs 1 and 2 are similar and the distribution is somewhat skewed to higher values.

The $W80$ and $W90$ parameters distributions -- shown in the third column of Figure\,\ref{fig:hist_w80_lowL}, present somewhat broader distributions for the QSOs 1 than for the QSOs 2, 
with the QSOs 2 presenting smaller median values for these parameters than type QSOs 1. 
Both populations present high velocity tails on the distributions that reach values of up to W80=2000\,\kms and W90=3000\,\kms, with these tails being stronger 
for type 1 QSOs. Finally, the $\langle$v$\rangle$ distribution -- shown in the fourth column of Figure\,\ref{fig:hist_w80_lowL}, is broader for the QSOs 1 and centred at bluer velocities, 
while for the QSOs 2 we find a narrower distribution, 
and centred closer to zero. The median values of each parameter, along with the respective KS test p-values, are presented in Table\,\ref{tab:median_kin_lowL}.

Due to the coupling of the kinematic components, the above results are also very similar for the \hb\ and \oiii$\lambda4363$ lines, with the latter only possessing the NLR$_\mathrm{n}$ component accounted for. The median values and KS test p-values for these lines are also included in Table\,\ref{tab:median_kin_lowL}, and differ only due to the minimum luminosity criterion adopted, in which we exclude the components with luminosities lower than 3 times the ``noise luminosity'', defined in Paper\,I as that of a Gaussian profile with amplitude equal to the spectral noise given by its RMS, and FWHM corresponding to the mean spectral resolution $\Delta\lambda = 1.25$\,\AA.

\medskip
\noindent{\bf \neiii\ and \nev\ --} Figure\,\ref{fig:hist_vnev_comp_lowL} shows in its panels, from top to bottom, the velocities and velocity dispersions of the \neiii\ and \nev\ emission lines, from left to right. As discussed in Paper\,I, these lines were fitted with only one component and are produced by ions with higher ionisation potentials than \oiii, and, therefore, trace even more strongly ionised gas in the galaxy.

The \neiii\ v distributions for QSOs 1 is somewhat more redhifted when compared to QSOs 2, with a redshifted median of $\sim$50\,\kms, while for QSOs 2 the median is $\sim$15\,\kms. For QSOs 1, the distribution is somewhat broader than for QSOs 2, reaching higher redshifted velocities above 250\,\kms, up to 500\,\kms. The $\sigma$ distribution show values in the range $\sim$100--500\,\kms, with type 1 QSOs presenting somewhat larger median than type 2's. The KS-test does not reveal any statistical similarities between QSOs 1 and 2 for the \neiii\ kinematic parameters.

For \nev, there is a bimodality in the velocity distributions: for QSOs 2, the profile is centred closer to zero, covering a smaller range of velocities than type 1 QSOs, for which the distribution shows an excess of blueshifts, centred at $\sim$-170\kms. The velocity dispersion distribution is also distinct between the two types: the type 2 QSO distribution presents smaller values than type 1 objects, with medians $\sim$215\,\kms and $\sim$360\,\kms, respectively. We also note that the $\sigma$ values for type 1 objects reach up to 500\,\kms, which was adopted as the upper limit for the velocity dispersion due to contamination by residuals of the fit of the \feii\ template. 

These results show that \nev\ traces a kinematically hotter region, present in QSOs 1, relative to that traced by the \neiii\ and QSOs 2 in general. The results from Paper\,I suggest that the \nev\ emitting gas is located closer to the ionization source and probing a harder ionisation continuum -- such as the radiation originating from the accretion disk -- than \neiii. The larger velocity dispersion in the former, along with its excess of blueshifts, supports this scenario. Therefore, it seems that the coronal line \nev\ is another potential indicator of outflows, tracing a more ionised gas phase.

In support of this scenario, we note that these lines are harder to detect for the LL sample, specially for type 2 sources, of which only 59 objects have \neiii\ detected and for \nev, only 36.

\subsection{Mass outflow rate and power}

\begin{figure*}
    \centering
    \includegraphics[width=0.96\textwidth]{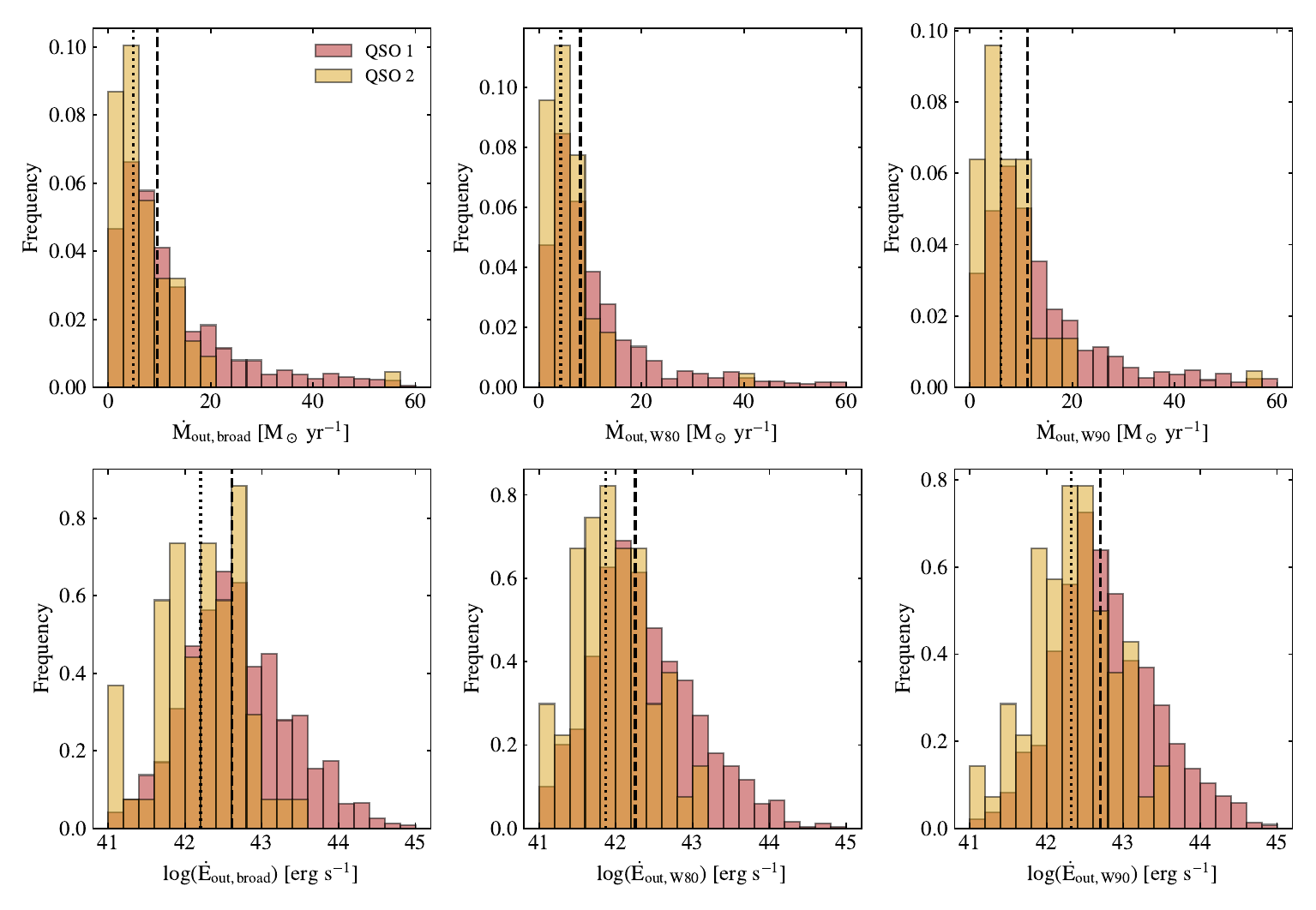}
    \caption{Normalized distributions of the mass outflow rate (top) and power (bottom) for the QSO 1 and QSO 2 populations, shown as red and yellow, respectively, for the LL sample. The left panels are calculated with the broad component method, the center with the W80 and the right, with the W90 method. The median of each population is shown as black dashed (for type 1) and dotted (for type 2) vertical black lines.}
    \label{fig:mdot_edot_lowL}
\end{figure*}

On the upper and lower panels of Figure\,\ref{fig:mdot_edot_lowL}, we present the mass outflow rates and the outflow kinetic power, respectively, for the three methodologies presented in Section\,\ref{sec:methods_outflow}. From left to right, we show the results for the outflow velocities assumed to be given by v$_b$, $W80$ and $W90$, respectively. The mean values are presented in Table\,\ref{tab:median_kin_lowL}.

With the broad component method, we obtain that the mass outflow rate and power distributions for QSOs 1 show a tail extending to higher values than for QSOs 2, reaching up to $\dot{\mathrm{M}}_\mathrm{out, W90} \sim$\,40\,\msun\,yr$^{-1}$, 
with a median of 9.6\,\msun\,yr$^{-1}$ for QSOs 1, while QSOs 2 reach typically lower values, 
with a median of 5\,\msun\,yr$^{-1}$. 
For $\dot{\mathrm{E}}_\mathrm{out, broad}$, QSOs 1 reach log\,($\dot{\mathrm{E}}_\mathrm{out, broad}) \approx 44$ (erg\,s$^{-1}$) with a median log\,($\dot{\mathrm{E}}_\mathrm{out, broad}) = 42.6$ (erg\,s$^{-1}$),
while QSOs 2 reach up to log\,($\dot{\mathrm{E}}_\mathrm{out, broad}) \approx 43$ (erg\,s$^{-1}$) with a median log\, ($\dot{\mathrm{E}}_\mathrm{out, broad}) = 42.2$ (erg\,s$^{-1}$),

Similar trends and values are found for the W80/W90 methods, with $\dot{\mathrm{M}}_\mathrm{out}$ median values for QSOs 1 being approximately twice those of QSOs 2, and $\dot{\mathrm{E}}_\mathrm{out}$ median values being approximately 0.5 dex higher.
Overall, the relative trends between the two quasar populations remain consistent: type 1 objects present stronger outflow signatures than type 2 sources. 

\section{Results - High Luminosity Sample}

In this section, we list the results regarding the HL subsample. First, we discuss the kinematic parameters obtained from the emission-line fits, then discuss the corresponding outflow mass-outflow rates and powers.

\subsection{NLR kinematic parameters from the line-profile fits} \label{sec:kin_res_highL}

\begin{figure*}
    \centering
    \includegraphics[width=0.8\textwidth]{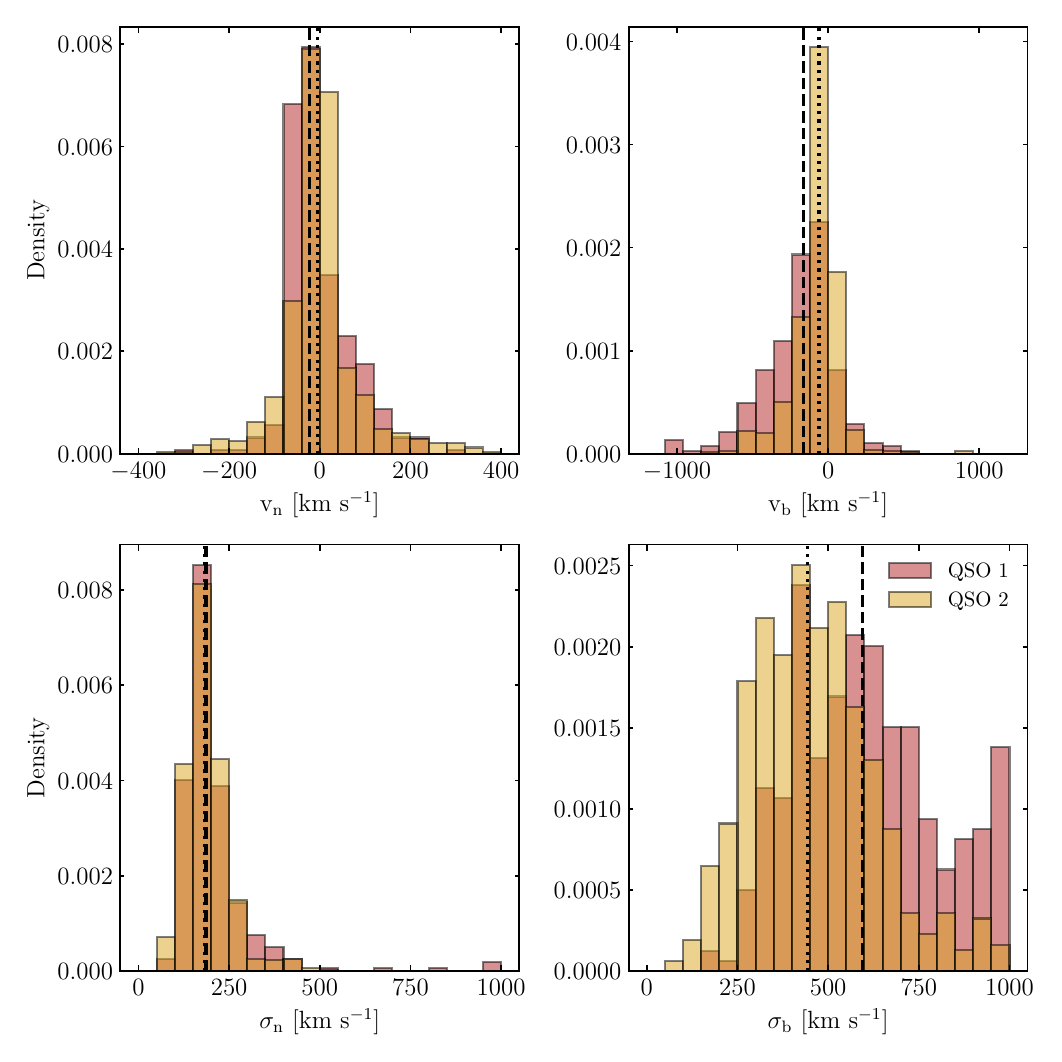}
    \caption{Normalized distributions of the velocity components v$_n$ and v$_b$ (top panels) and corresponding velocity dispersions $\sigma_n$ and $\sigma_b$ (bottom panels) for the \oiii$\lambda$5007 emission line, for the HL sample.  The QSO 1 and QSO 2 populations are shown as red and yellow histograms, respectively, with the median of each population shown as black dashed (QSO 1) or dotted (QSO 2) vertical black lines.}
    \label{fig:hist_voiii_comp_highL}
\end{figure*}

\begin{figure*}
    \centering
    \includegraphics[width=0.98\textwidth]{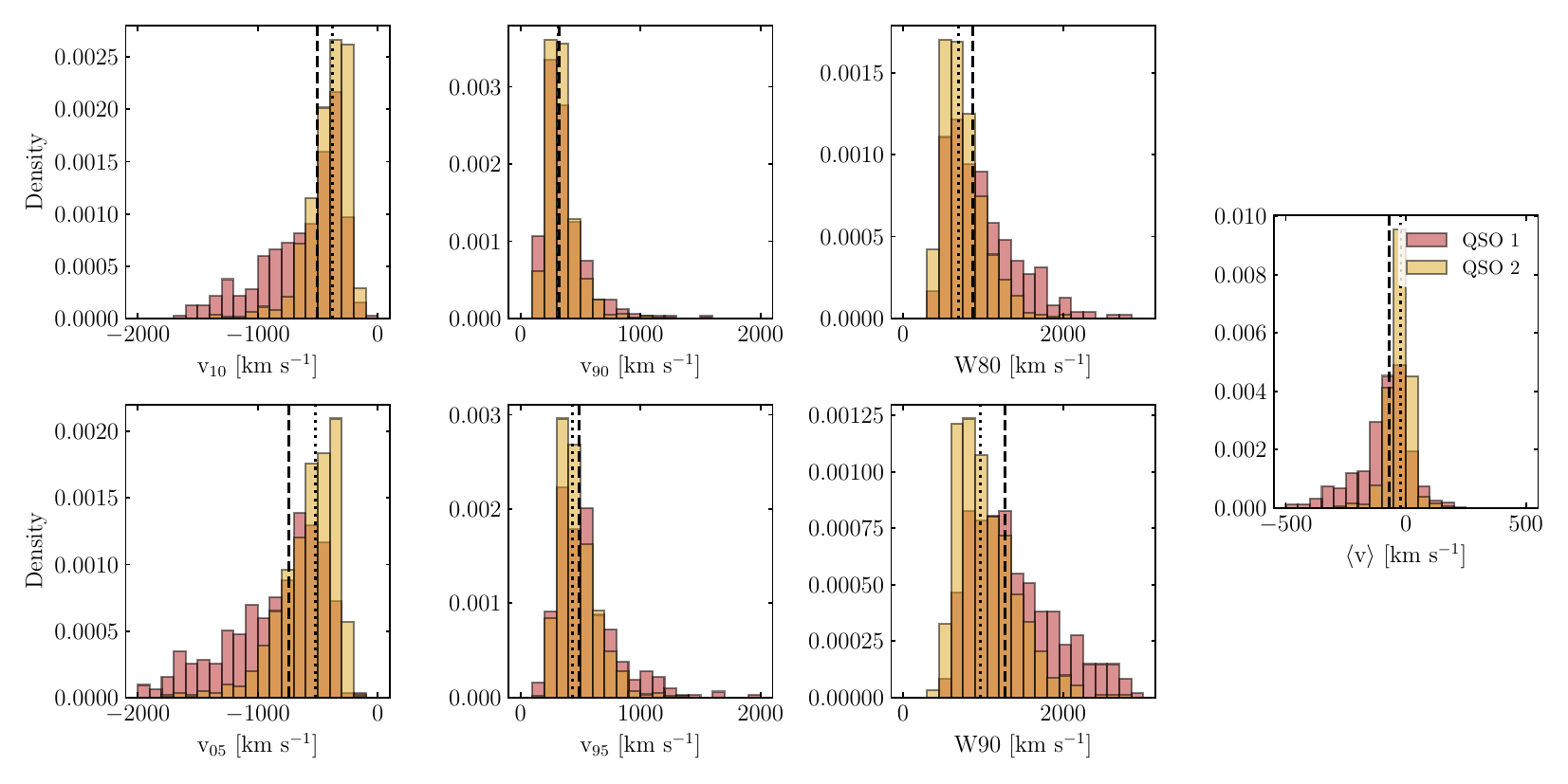}
    \caption{Normalized v$_{10}$, v$_{90}$, $W80$ (top row); v$_{05}$, v$_{95}$, W90 (bottom row) and $\langle$v$\rangle$ (right panel) distributions of the \oiii$\lambda$5007 emission line of the HL sample. The QSO 1 and QSO 2 populations are shown as red and yellow histograms, respectively, with the median of each population indicated as black dashed (QSOs 1) and dotted (QSOs 2) vertical black lines.}
    \label{fig:hist_w80_highL}
\end{figure*}

\begin{figure*}
    \centering
    \includegraphics[width=0.8\textwidth]{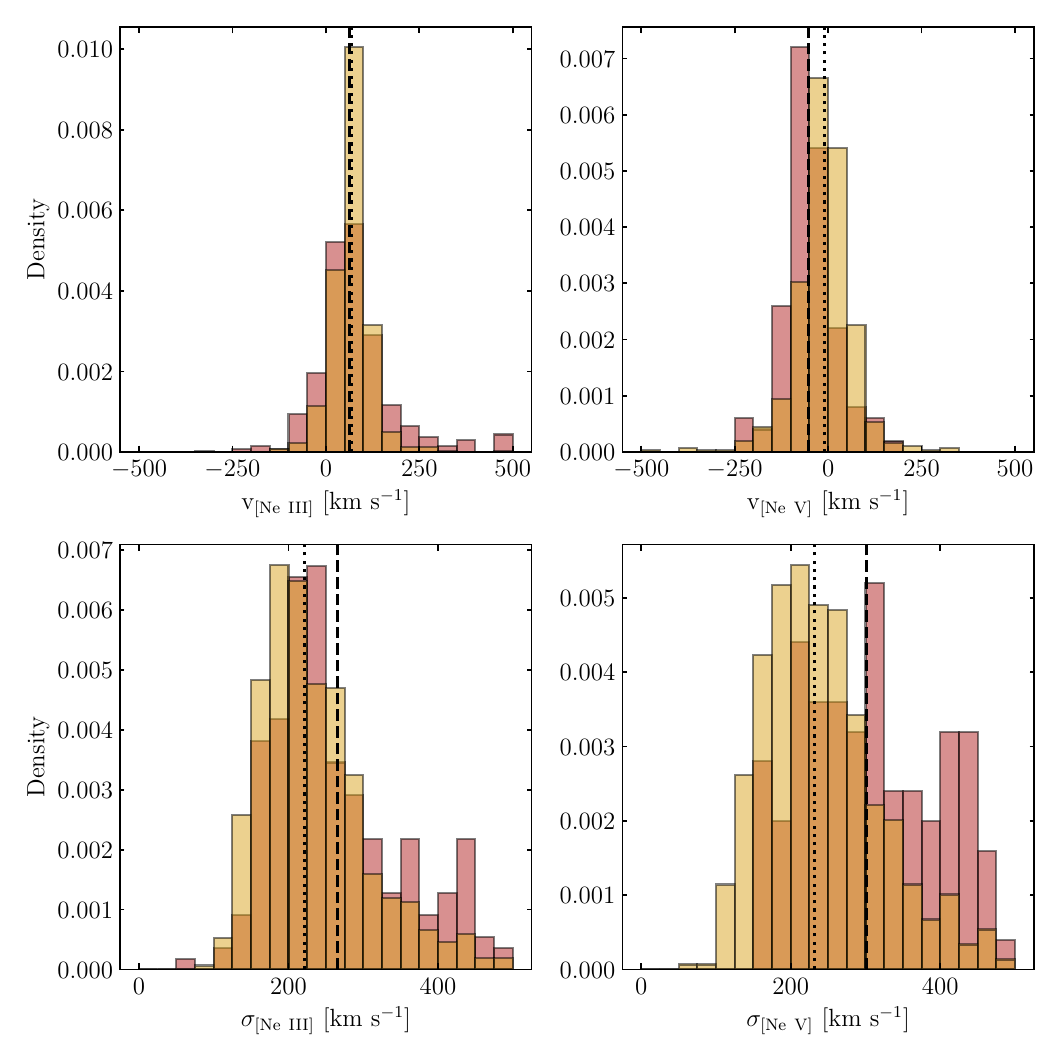}
    \caption{Normalized distributions of the velocity v and velocity dispersion $\sigma$ from the \neiii\ and \nev\ emission lines of the HL sample. The QSO 1 and QSO 2 populations are shown as red and yellow histograms, respectively, showing the corresponding median of each distribution as black dashed and dotted vertical black lines.}
    \label{fig:hist_vnev_comp_highL}
\end{figure*}

In this section we discuss the values of the parameters derived from the NLR emission-line profile fits: v$_\mathrm{n}$, $\sigma_n$, v$_ b$, $\sigma_b$, $W80$ and $W90$, comparing their distributions for the QSOs 1 and 2. The results are shown in Figs.\,\ref{fig:hist_voiii_comp_highL} -- \ref{fig:mdot_edot_highL} and the median values of the parameters and measurements are listed in Table \ref{tab:median_kin_highL}.

\medskip

\noindent{\bf\oiii$\lambda5007$ -- } Similarly to the analysis of the LL sample, in Figure\,\ref{fig:hist_voiii_comp_highL} we show, in the top panels, v$_n$ and v$_b$ and, in the bottom panels $\sigma_n$ and $\sigma_b$ obtained from the fit of the \oiii$\lambda$5007 emission-line for the two QSO types. 

The velocities v$_n$ and v$_b$ and corresponding velocity dispersion values $\sigma_n$ and $\sigma_b$ show very similar distributions to those of the LL sample: the v$_\mathrm{n}$ distributions are centred close to zero, while those for v$_\mathrm{b}$ are skewed towards negative values revealing more blueshifted broad components for type 1 QSOs. The velocity dispersion distributions, for both the narrow and broad components and type 1 and 2 QSOs, present a tail more extended towards higher values. The $\sigma_\mathrm{n}$ distribution again looks similar for both types, typically not reaching values larger than 500\,\kms, while the $\sigma_\mathrm{b}$ distribution reaches values up to 1000\,\kms\ for both populations but with more QSOs 1 presenting higher values. (The excess of values close to the limit 1000 km\,s$ ^{-1}$ in QSOs 1 is attributed to numerical degeneracy of the minimization algorithm used for fitting the spectra.) Of these panels, $\sigma_\mathrm{n}$ presents statistical similarity between QSOs 1 and 2, as revealed by a large p$_\mathrm{KS}$, equal to 0.28, suggesting that the narrow NLR component of both QSO types originates from similar regions (e.g., rotation in the galaxy gravitational potential), while $\sigma_\mathrm{b}$ is clearly displaced towards larger values for QSOs 1, indicating it is probing higher $\sigma$ regions in QSOs 1 than in QSOs 2. 

In the upper panels of Figure\,\ref{fig:hist_w80_highL}, we present the v$_{10}$, v$_{90}$ and $W80$ distributions, along with v$_{05}$, v$_{95}$ and $W90$ in the bottom panels, and $\langle$v$\rangle$ in a panel to the right, derived from the \oiii$\lambda$5007 profile. The individual panels are very similar to those for the LL sample in Figure\,\ref{fig:hist_w80_lowL}: v$_{05}$, v$_{10}$, v$_{90}$ and v$_{95}$ reveal that the profiles tend to be blueshifted, with QSOs 1 reaching up to twice the blueshifted values of v$_{05}$ and v$_{10}$ from QSOs 2. The values of v$_{90}$ and v$_{95}$ extend to similar redshifts; $W80$ and $W90$ show that type 1 \oiii$\lambda$5007 profiles are on average broader than type 2's; and the $\langle$v$\rangle$ distribution is broader for the QSOs 1 and centred at more negative velocities. 

We again mention that the above results are also very similar to those  for the \hb\ and \oiii$\lambda4363$ lines, due to the coupling of the kinematic components in the fitting procedure. The medians and KS test p-values for these lines are also included in Table\,\ref{tab:median_kin_highL}.

\medskip
\noindent{\bf \neiii\ and \nev\ --} The results regarding the \neiii\ and \nev\ emission lines kinematics are shown in the upper and lower panels of Figure\,\ref{fig:hist_vnev_comp_highL}, respectively. 

The \neiii\ velocity and $\sigma$ distributions are more similar between QSOs 1 and 2 than observed in the LL sample: (i) the emission lines are mildly redshifted with positive velocity medians and (ii) the $\sigma$ for QSOs 1 is somewhat larger than for QSOs 2, with the distribution showing values in the range $\sim$100--500\,\kms. But the KS statistics still indicate differences between the QSOs 1 and 2 distributions for both parameters.

For \nev, the apparent bimodality seen for the LL velocity distribution decreases, with QSOs 2 still possessing a symmetrical velocity distribution, centred close to zero, while QSOs 1 are still typically blueshifted, but at smaller (absolute) velocities (median $\sim$-50\,\kms). The bimodality previously found for the LL velocity dispersion is also weakened, but type 2 objects still present smaller $\sigma$ values than type 1's, with median $\sigma$ $\sim$300\,\kms for QSOs 1 and $\sim$230\,\kms\ for QSOs 2.

These results, again, supports the evidence that \nev\ emission is mostly located closer to the ionising source than \neiii\, and has a more disturbed kinematics (specially in QSOs 1) and likely traces outflowing material.

\subsection{Mass outflow rate and power}

\begin{figure*}
    \centering
    \includegraphics[width=0.96\textwidth]{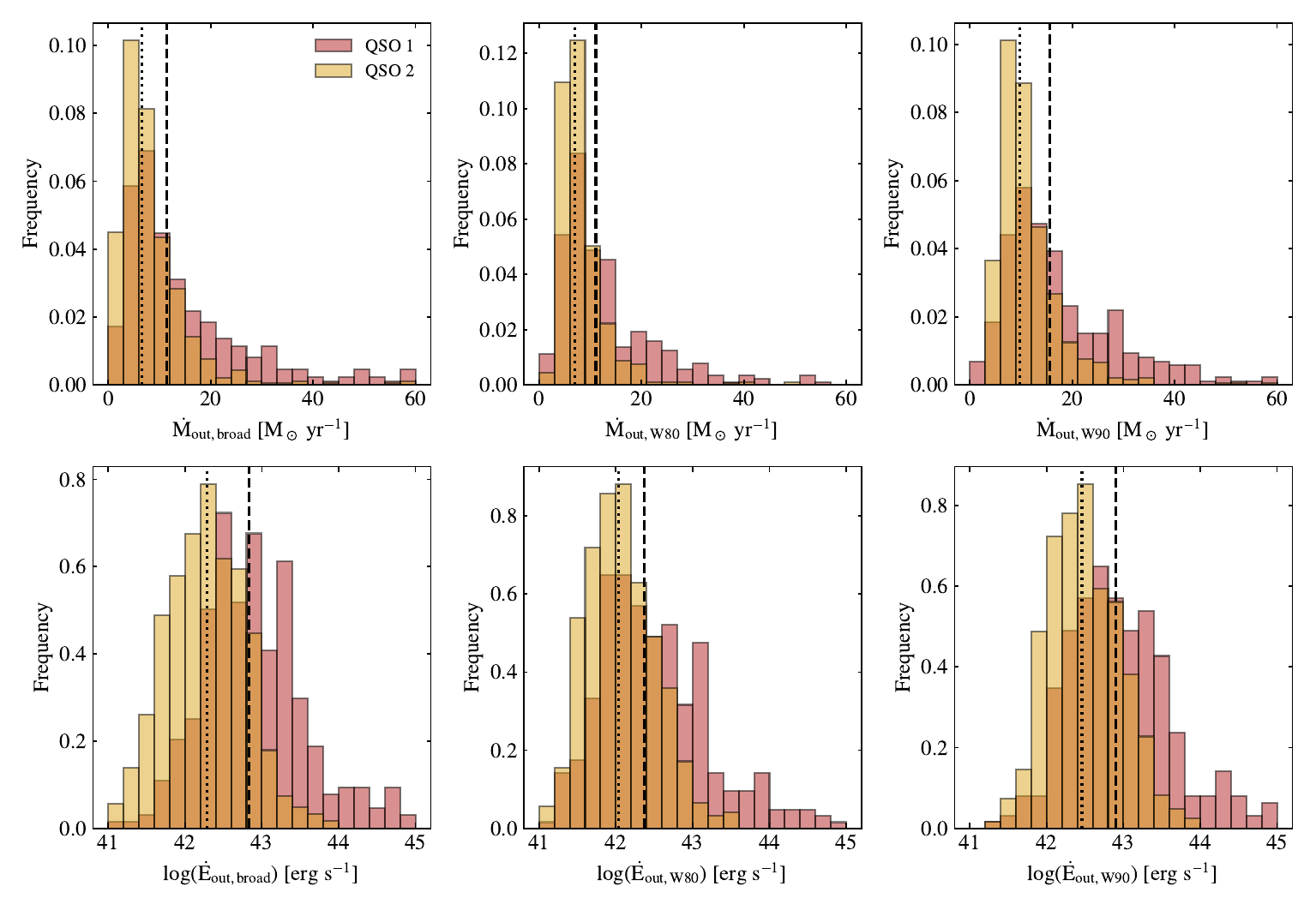}
    \caption{Normalized distributions of the mass outflow rate (top) and mass outflow power (bottom) for the QSO 1 and QSO 2 populations, shown as red and yellow, respectively, for the HL sample. The histograms in the left panels are calculated with the broad component method, the central ones with the W80 method and the right, with the W90 method. The median of each population is shown as black dashed or dotted vertical black lines.}
    \label{fig:mdot_edot_highL}
\end{figure*}

In the upper and lower panels of Figure\,\ref{fig:mdot_edot_highL} we show, respectively, the mass outflow rates and the kinetic power distributions for the HL sample. From left to right, we show the results obtained for the three methodologies presented in  Section\,\ref{sec:methods_outflow}: using the gas velocity as that of the broad component, and using the velocities as $W80$ and $W90$, respectively.

For the broad component method, we again find more extended $\dot{\mathrm{M}}_\mathrm{out, broad}$ and $\dot{\mathrm{E}}_\mathrm{out, broad}$ distributions, and also reaching higher values for QSOs 1 than for QSOs 2, up to log\,$\dot{\mathrm{M}}_\mathrm{out, broad}\approx\,45$ (erg\,s$^{-1}$) and $\approx\,44$ (erg\,s$^{-1}$), respectively.
The median values are: $\dot{\mathrm{M}}_\mathrm{out, broad} = 11.5$\,\msun\,yr$^{-1}$ and 6.6\,\msun\,yr$^{-1}$, respectively; and median log\,($\dot{\mathrm{E}}_\mathrm{out, broad}) = 42.8$ and 42.3, respectively.

Similar trends and values are found for the W80/W90 methods, with $\dot{\mathrm{M}}_\mathrm{out}$ median values for QSOs 1 being approximately 1.6 times higher than those of QSOs 2, and $\dot{\mathrm{E}}_\mathrm{out}$ median values being approximately 0.4--0.5 dex higher. Overall, the relative trends between the two quasar populations remain again consistent: type 1 objects present stronger outflow signatures than type 2 sources.


\section{Discussion}

We conclude that the three adopted velocity indicators for the outflows -- v$_b$, $W80$ and $W90$, give similar results for the kinematic measurements of $\dot{M}$ and $\dot{E}$, which are well correlated with each other, as illustrated in Fig.\,\ref{fig:scatterplot_kin_Lum}. The correlations between the adopted velocities is also shown in Fig.\,\ref{fig:scatterplot_W_vb}, supporting their use in studies of integrated kinematic properties of AGN, as in our study.
Due to these similarities, we adopt, in the following discussion, the broad component method, most widely used in the recent literature to trace outflows and for having been used in our reference paper of resolved NLR gas kinematics in QSOs \citet{dallagnol_de_oliveira_gauging_2021}.

\subsection{NLR kinematic parameters from the line-profile fits}
\label{NLR discussion}

From the histograms of Figures \ref{fig:hist_voiii_comp_lowL}--
\ref{fig:mdot_edot_highL}, as well as Tables \ref{tab:median_kin_lowL} and \ref{tab:median_kin_highL}, we find that:

\begin{itemize}

\item The median velocities of the \oiii\ narrow component v$_n$ are close to zero, for both QSOs 1 and 2, 
with similar distribution of values. The velocity dispersion $\sigma_n$ shows also similar median values (150--180\,km s$^{-1}$) 
and distributions.

\item These similar v$_n$ velocities  -- close to zero, and similarly low $\sigma_n$ 
indicates that the narrow component of the gas has similar origin in QSOs 1 and 2, most probably gas orbiting in the galaxy gravitational potential, ionized by the AGN. 

\item The velocities of the broad component v$_b$ are both negative but reach more negative values for QSOs 1 than 2, 
with similar median velocity dispersion values $\sigma_b$ in the range 450--600\,km\,s$^{-1}$.

\item Median $\sigma_\mathrm{b}$ values higher than typical circular velocities, together with blueshifted $v_\mathrm{b}$ values, suggest that the broad component serves as a signature of AGN kinetic feedback driven by outflows.

\item The excess blueshifts obtained for the QSOs 1 relative to the QSOs 2 can be attributed to the orientation of the outflow, consistent with that expected in the Unified Model scenario, in which QSOs 1 are observed more ``face-on", or ``pole-on", than the QSOs 2. The fact that the velocity is lower but still negative for the QSOs 2 can be due to possible extinction within the NLR, such that we see more easily the part of the outflow closer to us, that, in a conical outflow, is blueshifted. 

\item As a simple exercise to check the validity of the above hypothesis for the orientation of the outflow, if we consider that the AGN in QSOs 1 are approximately ``face-on", the angle $\theta$ between the orientation of the outflows in QSOs 2 relative to QSOs 1 can be estimated from the ratio between the median v$_b$ velocities of the two samples.
These ratios imply a difference in orientation of the outflows in QSOs 1 and 2 in the range 63$^\circ-68^\circ$. Considering typical angular half-apertures of AGN cones (along which the gas would be in outflow) of $30^\circ-45^\circ$, 63$^\circ-68^\circ$ differences in the orientation between QSOs 1 and 2 are consistent with the hypothesis that we are seeing the AGN in QSOs 1 more face-on and in the QSOs 2, more edge-on.

\item The median values and distributions of the blueshift velocities v$_{05}$ and v$_{10}$ from \oiii\ also support the interpretation above: they show more negative median values for the QSOs 1 (up to -811\,km\,s$^{-1}$) 
than for QSOs 2 (up to -612\,km\,s$^{-1}$).

\item The $W80$ and $W90$ \oiii\ median values are consistent with the v$_b$ and $\sigma_b$ values, also supporting the more face-on orientation for the QSOs 1, that show larger profile widths than QSOs 2. 

\item Among the other emission lines, fitted with only one Gaussian component, the most significant difference between QSOs 1 and 2 was found for the coronal line of \nev: although both show blueshifted median values, QSOs 1 show an excess blueshift and higher
median velocity dispersion than QSOs 2.

\item The coronal line \nev\ kinematics suggests that we are observing the emitting gas closer to the nucleus, where outflows are more energetic, the temperature is higher, with the presence of shocks. This result can also be attributed to the effect of inclination, where the outflows are more pole-on in the QSOs 1 than in QSOs 2.

\end{itemize}

In summary, both type 1 and type 2 QSOs present a blueshifted NLR broad component, reaching higher velocities in QSOs 1. The broadening and blue asymmetry is a well-established indicator of non-circular motions, commonly interpreted as outflows emerging from the AGN. The ubiquity of this feature in our large quasar sample supports that ionised gas outflows are a fundamental component of AGN activity \citep[e.g.][]{harrison_agn_2018}.

This scenario aligns with previous observational evidence across a broad redshift and AGN luminosities range. \citet{komossa_nature_2008, mullaney_narrow-line_2013} showed that \oiii\ profiles frequently display asymmetries and velocity offsets, with the blueshifted wing being a typical signature of outflows. \citet{zakamska_quasar_2014, zakamska_discovery_2016} further confirmed that these features are prevalent among obscured (type 2) QSOs, particularly in systems with powerful feedback signatures such as extended radio emission or high Eddington ratios, up to $z = 2.5$. More recently, \citet{coatman_kinematics_2019} and \citet{deconto-machado_exploring_2024} demonstrated that \oiii\ blueshifted wings are not only common but increase in strength with AGN luminosity and redshift, with up to 60\% of quasars showing 
broad components attributed to outflows.

The prevalence of a broad and mostly blueshifted component in \oiii\ in most AGN  supports the interpretation of AGN-driven outflows that propagate through the host galaxy's ISM, potentially regulating star formation through feedback mechanisms \citep[e.g.,][]{harrison_agn_2018, audibert_jet-induced_2023, harrison_observational_2024}. 

Single-fiber spectroscopic studies, like ours, provide only statistical evidence for these kinematic signatures, but spatially resolved studies with IFS data have been offering more direct confirmation of the presence of outflows by spatially resolving the structure and extent of these outflows, which can reach kpc scales \citep{harrison_kiloparsec-scale_2014, dallagnol_de_oliveira_gauging_2021, gatto_extent_2024, revalski_quantifying_2022}.

The findings of blue wings also in the coronal line of \nev\ align with previous observational studies of such lines \citep{rodriguez-ardila_outflows_2006, rodriguez-ardila_near-infrared_2011, rodriguez-ardila_coronal_2025, rodriguez-ardila_size_2025}, which find evidence that these coronal lines are concentrated at the innermost regions of the NLR and show evidence of shocks and outflows.

\subsection{Comparison with the NLR excitation from Paper I}

In \citetalias{hauschild-roier_differences_2025}, we have shown that QSOs 1 present signatures of shock excitation as derived from the high values of \nev/\neiii\ and \oiii\,4364/5007, indicating higher electronic temperatures  -- up to 50000\,K, which is inconsistent with pure radiative excitation \citep[e.g.][]{dopita_spectral_1995, osterbrock_astrophysics_2006, allen_mappings_2008}. Most of the QSO 2 population present lower electronic temperatures and emission line ratio distributions with medians around values consistent with radiative excitation, including higher \oiii/H$\beta$ ratios in QSOs 2. Therefore, we concluded that QSOs 1 NLR excitation has a higher contribution from shocks, while the radiation from the accretion disk alone is the most likely mechanism to produce the excitation of the NLR in the QSOs 2. 

This result is consistent with the kinematics: the higher blueshift emission lines found for QSOs 1, along with the larger $W80$ and $W90$ reached in the NLR indicate higher velocities  for the outflowing gas, consistent with a larger contribution from shocks. 

As pointed out in the previous section, there is evidence (from the median ratios of v$_b$ between the QSOs 2 and 1) that the higher temperatures and shock signatures in QSOs 1 can be attributed to the effect of inclination: the QSOs 1 are observed more pole-on than QSOs 2, allowing the observation of gas closer to the AGN, reaching regions where the gas outflow velocities and temperatures are higher than in QSOs 2. 

But we cannot discard some additional contribution from evolution, such that type 2 QSOs represent an earlier, dust-enshrouded phase of black hole growth, during which obscuring material has not yet been cleared by AGN feedback. As the outflow and AGN radiation clears the circumnuclear medium, the system evolves into a type 1 QSO, where the central engine becomes directly visible as well as the outflows closer to the nucleus \citep[e.g.,][]{hopkins_black_2005, glikman_first-2mass_2012}. Evidences for interactions have indeed been observed in previous studies of the environment of the QSOs 2 of the Reyes subsample, showing an excess of mergers In QSOs 2 relative to a control sample \citep{araujo_nuclear_2023,pierce_galaxy_2023}, as well as in other samples \citep[e.g.[]{tozzi_super_2024}. In addition, as shown in \citetalias{hauschild-roier_differences_2025}, the QSOs 2 have higher \oiii/H$\beta$ ratios with L\oiii\ distributions peaking at higher luminosities than QSOs 1, as shown also in Fig.\,\ref{fig:Loiii_distribution} that we have included in the Appendix. This indicates the presence, on average, of a larger reservoir of gas ionized by the AGN in QSOs 2, most probably originating from these interactions. The presence of interactions is clearly observed in the images and spectroscopy of the QSO 2 sample in \citet{storchi-bergmann_bipolar_2018,dallagnol_de_oliveira_gauging_2021}.

\subsection{Relation between kinematics and luminosity}

Comparing the results from Tables \ref{tab:median_kin_lowL} and \ref{tab:median_kin_highL}, the HL sample has a luminosity $\approx 0.6$\,dex larger than that of the LL sample, with the corresponding median $\dot{\mathrm{M}}_\mathrm{out}$ and $\dot{\mathrm{E}}_\mathrm{out}$ values also increasing, but only by $\approx$\,20\%\ in the former and 0.2\,dex in the latter. Thus, although there is some effect of luminosity, it is small, that can be attributed, at least in part, to the small range of redshifts of our sample (0.4--0.5) and the Malmquist bias \citep{bigot_comments_1990,butkevich_statistical_2005}. This bias makes the sample to be restricted to a narrow range of luminosities. The effect of the scatter/uncertainties in kinematic properties seems to be larger than the effect of the range of luminosities of our sample.

Fig.\,\ref{fig:scatterplot_kin_Lum} of the Appendix shows that this seems indeed to be the case: there are no significant correlations between our various kinematic measurements and the AGN luminosity. But, in these scatter plots, we have identified the two QSOs 2 (Reyes and Yuan) samples using different colors. This allows to verify that the some distributions of properties differ between the Reyes and Yuan, in particular $W80$, $W90$ and v$_{out}$, that peak at lower values for the Yuan sample than for the Reyes sample; the latter sample shows similar distributions to those of QSOs 1. 


\subsection{Mass outflow rates and powers}

Figs.\,\ref{fig:mdot_edot_lowL} and \ref{fig:mdot_edot_highL} and Tables \ref{tab:median_kin_lowL} and \ref{tab:median_kin_highL} show that the median values derived for $\dot{\mathrm{M}}_\mathrm{out}$, for the three methods, vary at most by a factor of 1.5 and those for $\dot{\mathrm{E}}_\mathrm{out}$ by at most 0.5 dex, indicating that they give mostly equivalent values. Fig. \ref{fig:scatterplot_diff_methods} of the Appendix also shows that the three adopted outflow velocities (v$_b$, $W80$ and $W90$), result in very similar and correlated $\dot{\mathrm{M}}_\mathrm{out}$ and $\dot{\mathrm{E}}_\mathrm{out}$ values, supporting their use as similar indicators of kinetic feedback in AGN.

As $\mathrm{v_{out,broad}}$ provides similar $\dot{\mathrm{M}}_\mathrm{out}$ and $\dot{\mathrm{E}}_\mathrm{out}$ values to those of the other methods, and is more clearly identified with an outflow velocity, we will use it as our outflow tracer in the following discussion, where we compare our results with those from previous studies in the literature, adopting the simplified notation $\dot{\mathrm{M}}_\mathrm{out} = \dot{\mathrm{M}}_\mathrm{out, broad}$ and $\dot{\mathrm{E}}_\mathrm{out} = \dot{\mathrm{E}}_\mathrm{out, broad}$ from here on.
In addition, in the resolved study of \citet{dallagnol_de_oliveira_gauging_2021}, the uncertainties in the values of $\dot{\mathrm{M}}_\mathrm{out}$ and $\dot{\mathrm{E}}_\mathrm{out}$ are larger than the differences among the median values of these quantities obtained using the three different methods.

\subsubsection{Coupling efficiency}

In order to investigate how our estimated outflow properties are coupled to the AGN luminosity L$_{bol}$, we plot $\dot{\mathrm{M}}_\mathrm{out}$ and $\dot{\mathrm{E}}_\mathrm{out}$ as a function of L$_\mathrm{bol}$, for both the LL and HL samples in Figure\,\ref{fig:mdot_edot_lbol}. Under the assumption that the lower negative velocity v$_b$ in QSOs 2 than in QSOs 1 is due to its higher inclination, we have divided the QSOs 2 v$_b$ values by cos(65$^\circ$), for a mean difference in inclination of 65$^\circ$, to correct for the effect of inclination.

We also plot, for comparison, previous similar data from the literature \citep{fiore_agn_2017, baron_discovering_2019, davies_ionized_2020, dallagnol_de_oliveira_gauging_2021, revalski_quantifying_2021, trindade_falcao_hubble_2021, bianchin_gemini_2022, gatto_extent_2024}.

According to models of AGN feedback, in order to affect the galaxy, delaying or halting star formation, the feedback power has to achieve a certain threshold in coupling efficiency $\varepsilon=\dot{\mathrm{E}}/\mathrm{L}_\mathrm{bol}$. In order to compare our data with such models, we also include, in the right panel of Figure\,\ref{fig:mdot_edot_lbol}, threshold lines from AGN feedback models \citep{di_matteo_energy_2005, hopkins_quasar_2010} as well as simulations, such as Illustris-TNG \citep{weinberger_simulating_2017} and EAGLE. We have calculated also the average coupling efficiency, obtained from our data points, separately for the QSOs 1 (red) and 2 (yellow). 

Figure\,\ref{fig:mdot_edot_lbol} shows that our QSO 1 sample reaches somewhat higher $\dot{\mathrm{M}}_\mathrm{out}$ (left panel) and $\dot{\mathrm{E}}_\mathrm{out}$ (right panel) values than the QSOs 2, with these latter being consistent with the values of the reference paper of \citet{dallagnol_de_oliveira_gauging_2021}, as well as those of \citet{trindade_falcao_hubble_2021,revalski_quantifying_2021, fiore_agn_2017}. The average coupling efficiency for QSOs 1 is larger than for QSOs 2, with $\varepsilon_\mathrm{QSO\ 1} = 0.26 \%$ and $\varepsilon_\mathrm{QSO\ 2} = 0.04 \%$, respectively, but with large scatter.

Our estimates thus suggest that QSOs 1 produce a stronger kinetic feedback on its host galaxy than QSOs 2, with $\sim$30\%\ sources laying above the theoretical minimum efficiency line of \citet{hopkins_quasar_2010}, suggesting that kinetic feedback from QSOs 1 could suppress star formation in their host galaxies, according to models. Most QSOs 2 are below QSOs 1 and the models thresholds, but show similar coupling efficienties to that obtained from previous studies for AGN of the same L$_\mathrm{bol}$.




\subsubsection{Uncertainties}
\label{uncertainties}

Sources of uncertainty in the $\dot{\mathrm{M}}_\mathrm{out}$ and $\dot{\mathrm{E}}_\mathrm{out}$ mostly comprise the gas density, velocity, spatial distribution and extinction.

\smallskip

\noindent {\it Regarding the gas density}: We adopted a density of $\sim$\,300\,cm$^{-3}$ for the outflowing gas determined by \citet{dallagnol_de_oliveira_gauging_2021} in their resolved study of similar QSOs 2 from the \sii\ line ratio. But these same authors, as well as others \citep[e.g.][]{baron_discovering_2019, davies_ionized_2020,revalski_quantifying_2022} point out that the outflow density is most probably higher,  $\approx$\,1000\,cm$^{-3}$, leading to a value of $\dot{\mathrm{M}}_\mathrm{out}$ and $\dot{\mathrm{E}}_\mathrm{out}$ $\sim$ 3 times lower.

\smallskip

\noindent {\it Regarding the outflow velocity}: We adopted v$_b$+2$\sigma_b$, but, adopting instead a smaller velocity of v$_b$+$\sigma_b$ \citep[e.g.][]{dallagnol_de_oliveira_gauging_2021}, the results show, on average, a decrease by a factor of $\sim$\,2 for $\dot{\mathrm{M}}_\mathrm{out}$ and of $\sim$\,5 for $\dot{\mathrm{E}}_\mathrm{out}$.

\smallskip

\noindent {\it Regarding the geometry and orientation of the outflow}: If we considered, for example, a spherical outflow, we would obtain $\approx$\,3 times larger $\dot{\mathrm{M}}_\mathrm{out}$ and $\dot{\mathrm{E}}_\mathrm{out}$ values. Regarding the orientation, we derived, from the ratio of the two median v$_b$ velocities, a difference in inclination between the QSOs 2 and QSOs 1 of 65$^\circ$, and have corrected v$_b$ accordingly for the QSOs 2, to minimize the effect of different orientations of the outflow.

\smallskip

\noindent {\it Regarding the reddening:} We have adopted a median reddening -- based on the resolved study of \citet{dallagnol_de_oliveira_gauging_2021}, as well as on the previous studies of \citet{reyes_space_2008} and \citet{lu_reddening_2019}, of $E(B-V)=0.4$, with an estimated uncertainty of $E(B-V)=0.2$, corresponding to change by a factor of $\approx 2$ in $\dot{M}$ or a difference of $\approx 0.3$ in log\,$\dot{M}$, log\,$\dot{E}$ and log\,L$_{bol}$.

\begin{figure*}
    \centering
    \includegraphics[width=0.99\textwidth]{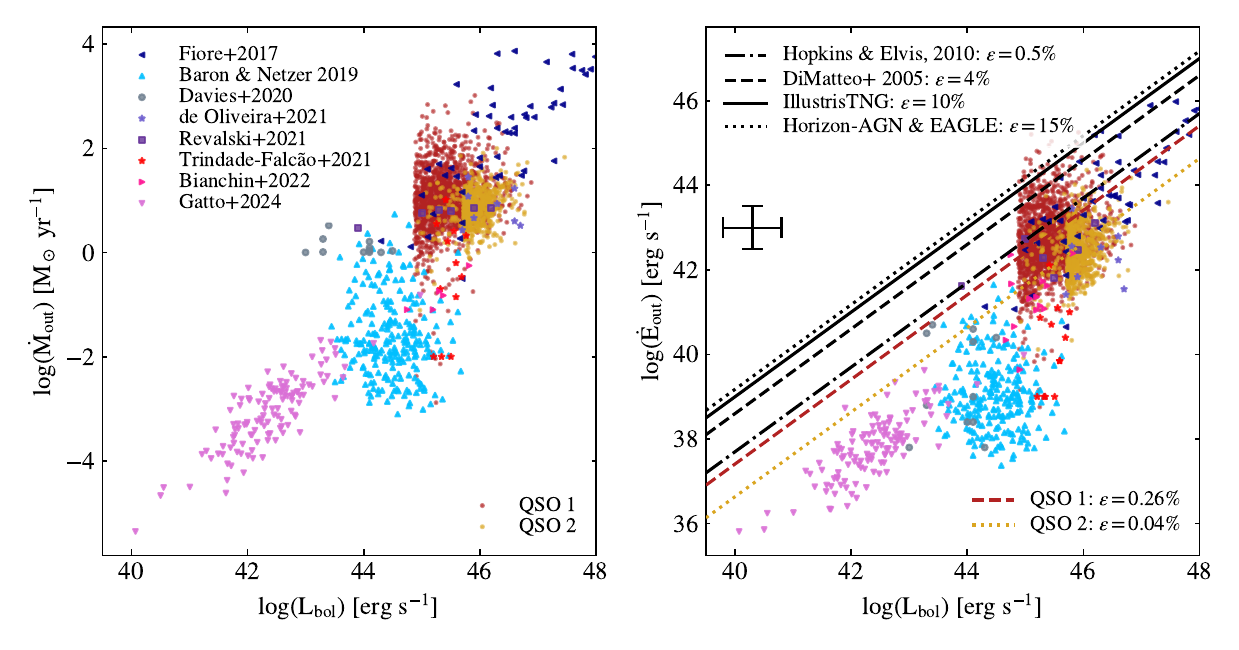}
    \caption{$\dot{\mathrm{M}}_\mathrm{out}$--L$_\mathrm{bol}$ and $\dot{\mathrm{E}}_\mathrm{out}$--L$_\mathrm{bol}$ planes derived with the broad component method, where the QSOs 2 v$_b$ values have been corrected for inclination. 
    For comparison, we add other data points from the literature and also include the accretion efficiency $\varepsilon$ lines from models. We include the QSOs 1 and 2 as red and yellow points, respectively, and show also, as dashed lines, the best fit $\varepsilon$ lines for them at their respective colours.}
    \label{fig:mdot_edot_lbol}
\end{figure*}

In summary, we estimate that the uncertainties in the parameters discussed above lead to mean uncertainties of $\sim 0.25$\,dex in log\,$\dot{\mathrm{M}}_\mathrm{out}$ and $\sim 0.5$\,dex in log\,$\dot{\mathrm{E}}_\mathrm{out}$ and log\,$L_{bol}$. We have included corresponding error bars in Fig.\,\ref{fig:mdot_edot_lbol}.

\section{Summary and Conclusions}

In this work, we have analysed the integrated  ionised gas kinematic properties from the Narrow-Line Region (NLR) of a SDSS spectroscopic sample of 1541 QSOs 1 and 688 QSOs 2 at redshifts $0.4 \leq z \leq 0.5$, comparing the results for QSOs 1 and 2. These integrated properties -- obtained within apertures at the galaxies of $\sim$\,8--9\,kpc radius -- comprise velocities, velocity dispersion values, $W80$ and $W90$ of the strongest NLR emission lines. In the case of H$\beta$ and \oiii$\lambda$5007, the profiles were fitted with two components: a narrow and a broad. 
As done in \citetalias{hauschild-roier_differences_2025} we subdivide the sample into: a low luminosity (LL) subsample, comprising objects with L$_{\mathrm{\oiii}}< 10^9$\lsun; and a high luminosity (HL) one, for objects with L$_{\mathrm{\oiii}}> 10^9$\lsun. 

Rather than a precise measurement of spatially resolved outflows, our analysis of a relatively large sample provides an statistical comparison of the NLR global kinematics between QSOs 1 and 2. Our main findings are:  

\begin{itemize}
    \item The kinematics of the narrow component of \oiii\ and H$\beta$ -- traced by v$_n$ and $\sigma_n$ -- is similar for QSOs 1 and 2, for the LL and HL subsamples, with median v$_n$ values close to zero and $\sigma_n$ values below $\sim$\,200\,km\,s$^{-1}$. These properties support the interpretation that the narrow component traces gas orbiting in the gravitational potential of its host galaxy;
    
    \item The kinematics of the broad component -- traced by v$_b$ and $\sigma_b$ -- shows instead stronger blueshifts in v$_b$ for QSOs 1, with $\sigma_\mathrm{b}$ being similar to those of QSOs 2 for the LL sample but $\approx$\,30\% broader for the HL sample. These properties suggest that v$_b$ traces outflowing gas, both in QSOs 1 and 2, but probing higher velocities and more disturbed gas in QSOs 1; 

    \item The kinematics of the broad component can also be traced by the observed larger values of $W80$, $W90$ and more blueshifted values for v$_{05}$ and v$_{10}$ in QSOs 1 than in QSOs 2;

    \item Another tracer of distinct NLR kinematics between QSOs 1 and 2 is the coronal line of  \nev, displaying blueshifted centroids, more so in QSOs 1, that show also larger velocity dispersions. This is consistent with an origin from inner, hotter and shock-excited regions, closer to the nucleus in QSOs 1 than in QSOs 2, as concluded in the study of the gas excitation in \citetalias{hauschild-roier_differences_2025}; 

    \item To estimate the AGN kinetic feedback and compare it with literature values, we derived approximate mass outflow rates ($\dot{M}_\mathrm{out}$), kinetic powers ($\dot{E}_\mathrm{out}$), and coupling efficiencies ($\varepsilon = \dot{E}_\mathrm{out} / L_\mathrm{bol}$). We find that QSOs 1 exhibit median $\dot{M}_\mathrm{out}$ values roughly twice those of QSOs 2, and median $\dot{E}_\mathrm{out}$ values approximately $0.5\text{ dex}$ higher. Furthermore, a global fit yields coupling efficiencies ($\varepsilon$) for QSOs 1 that are $1\text{ dex}$ above those for QSOs 2.

    \item When compared to previous results from the literature, the QSOs 2 $\dot{\mathrm{E}}_\mathrm{out}$ values are similar to those of other AGN of the same luminosities, while $\sim$ 30\%\ of the QSOs 1 show higher values; when compared to models, most QSOs 2 show $\varepsilon$ values below the minimum 0.05\% threshold for effective feedback on the galaxy, while for 30\%\ of the QSOs 1 these values are above this minimum.

\end{itemize}



Although our comparison between the QSOs 1 and 2 properties is robust, the inferred energetics depend on uncertain physical parameters of the NLR such as the gas density, geometry, radius, extinction, and projection assumptions, that cannot be resolved in the integrated measurements we have. On the other hand, the large number of sources provide statistical estimates that support the following conclusions:

(i) The differences between QSOs 1 and 2 can be understood as largely due to a more ``pole-on"  orientation for QSOs 1 than QSOs 2 relative to the line-of-sight, in support of the Unified Model. The result is an increased contribution to the gas emission from inner and hotter regions in QSOs 1 than in QSOs 2, characterized also by higher velocities and velocity dispersions. 

(ii) These differences can also be partially due to evolutionary effects, in which QSOs 2 represent an earlier, dust-enshrouded phase of black hole growth. As the outflow expands due to both shocks and radiation pressure and blow away the circumnuclear medium, the system could evolve into a type 1 QSO, where the central engine would become directly visible and the AGN outflows be better observed. The evolutionary scenario is also supported by the \oiii\ luminosity distribution of the QSOs 2 and QSOs 1 samples, that peak at higher L\oiii\ for the QSOs 2 than for QSOs 1, suggesting the presence of larger reservoirs of gas ionised by the AGN in QSOs 2.


\section*{Acknowledgements}

The authors thank an anonymous referee for raising a number of relevant points that led to a much improved manuscript. TSB acknowledges the brazilian Conselho Nacional de Desenvolvimento Científico e Tecnológico (CNPq) for the support for this research (Proj. CNPq-309928/2023-0). RR also acknowledges support from CNPq (Projs. 311223/2020-6, 304927/2017-1, 400352/2016-8, and 404238/2021-1), Funda\c{c}\~ao de Amparo \`{a} Pesquisa do Rio Grande do Sul (FAPERGS, Proj. 19/1750-2 and 24/2551-0001282-6) and Coordena\c{c}\~ao de Aperfei\c{c}oamento de Pessoal de N\'{i}vel Superior (CAPES, Proj. 0001). 

\section*{Data Availability}

The spectra used in this work are available for download at the SDSS-IV DR16 website (https://www.sdss4.org/dr16/). All figures produced from these data, and the data itself, can be shared on reasonable request to the corresponding author.



\bibliographystyle{mnras}
\bibliography{references}



\appendix

\section{Relation between kinematic properties and the AGN luminosity}

Here we present a few scatter plots showing the relation between kinematic parameters obtained for the three gas velocity indicators (v$_b$, $W80$ and $W90$) as well as the relation between the kinematic parameters and the AGN luminosity, highlighting the three individual samples used in this study. We also include the Kernel Density Estimation (KDE) for each parameter for each sample instead of histograms, due to the different sizes of the samples. 

These plots do not show significant variations of the parameters with luminosity, most probably due to the fact that the range in luminosities is small, bellow 2 dex. These plots also show the range of the kinematic properties values of our sample (whose medians are shown in tables \ref{tab:median_kin_lowL} and \ref{tab:median_kin_highL}).

\begin{figure*}
    \centering
    \includegraphics[width=0.70\textwidth]{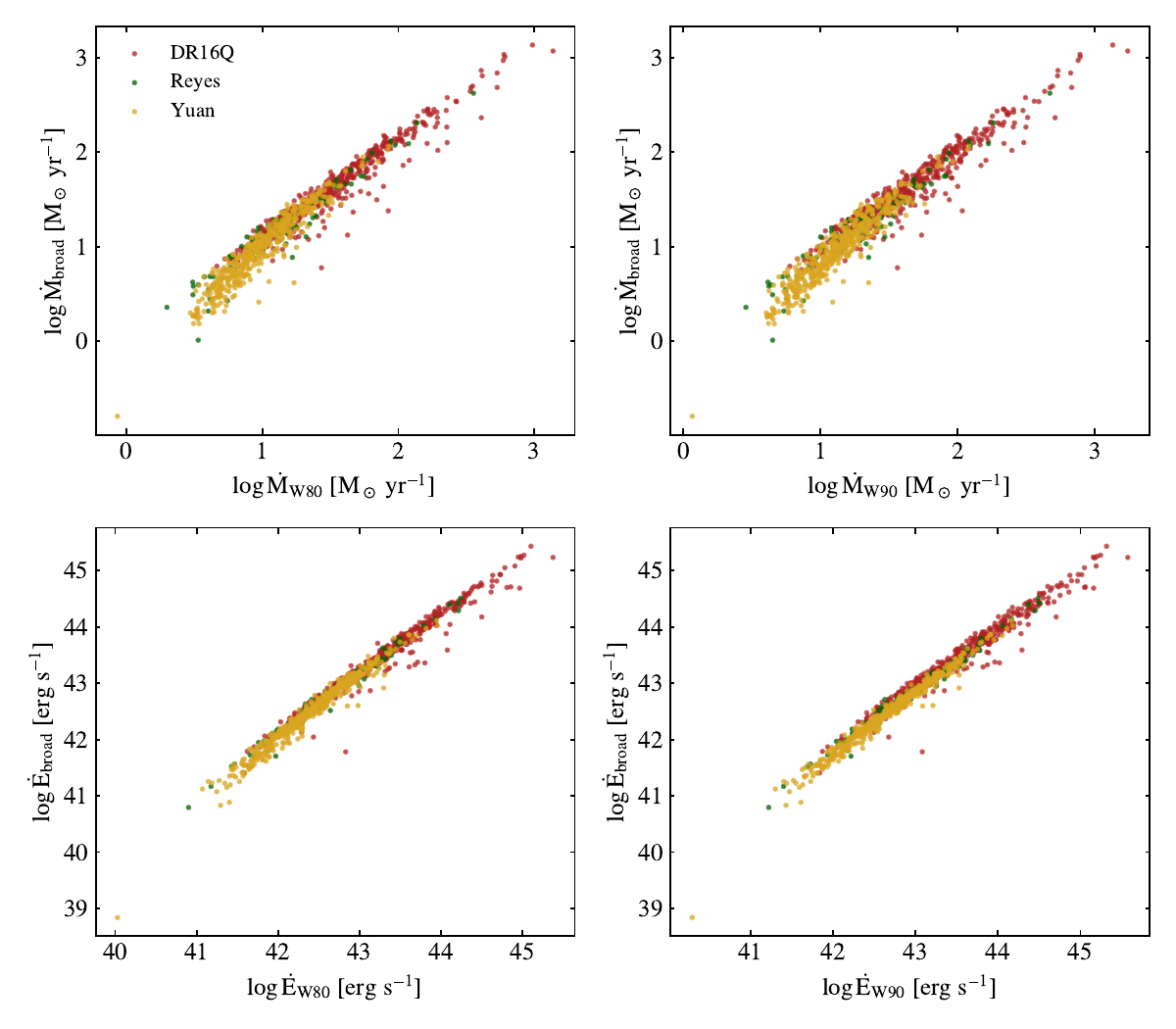}
    \caption{Relation between the measured $\dot{\mathrm{M}}_\mathrm{out}$ and $\dot{\mathrm{E}}_\mathrm{out}$ using the three different velocities: of the broad component, $W80$ and $W90$. The QSOs 1 are shown as red dots, the QSOs 2 from the Reyes sample are shown in green and those from the Yuan sample are shown in yellow.}
    \label{fig:scatterplot_diff_methods}
\end{figure*}

\begin{figure*}
    \centering
    \includegraphics[width=0.96\textwidth]{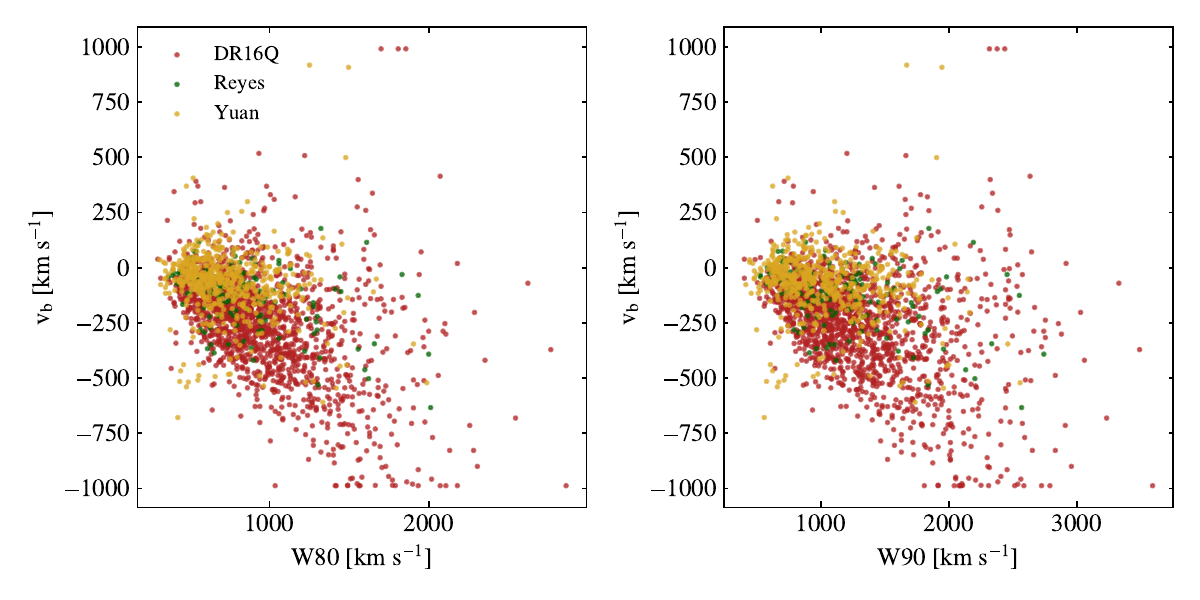}
    \caption{Relation between the measured v$_\mathrm{b}$, W80 and W90 of the QSOs 1 and 2. The QSOs 1 are shown as red dots, the QSOs 2 from the Reyes sample are shown in green and those from the Yuan sample are shown in yellow.}
    \label{fig:scatterplot_W_vb}
\end{figure*}

\begin{figure*}
    \centering
    \includegraphics[width=0.56\textwidth]{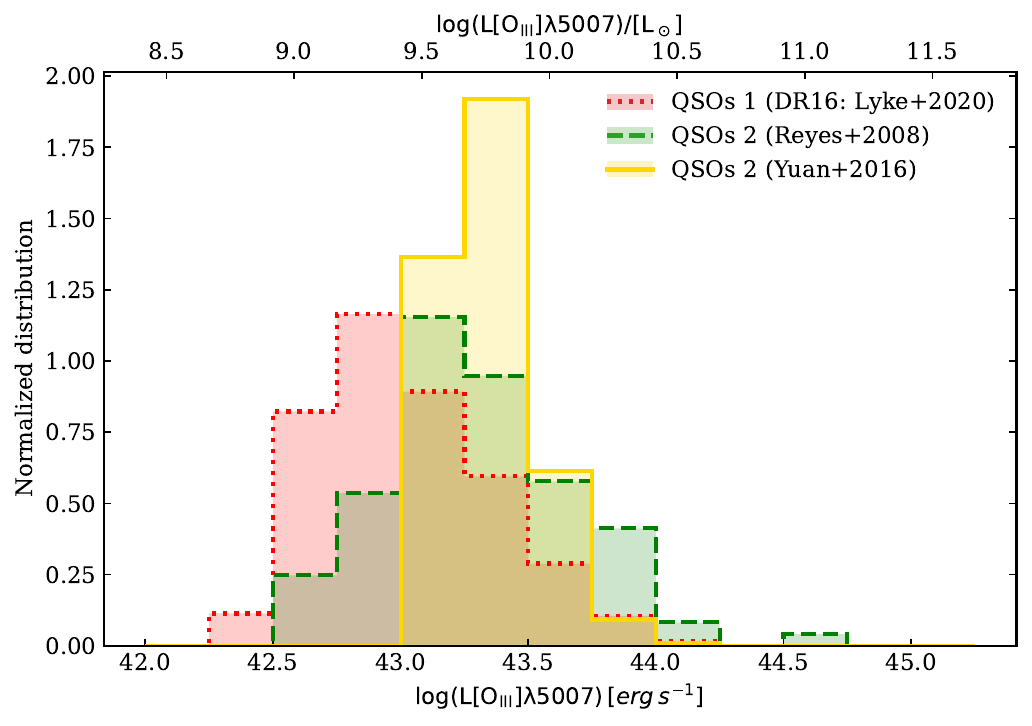}
    \caption{L\oiii\ normalized distributions for the QSOs 1 and the two samples of QSOs 2. The QSOs 1 are shown as red dots, the QSOs 2 from the Reyes sample are shown in green and those from the Yuan sample are shown in yellow.}
    \label{fig:Loiii_distribution}
\end{figure*}

\begin{figure*}
    \centering
    \includegraphics[width=0.96\textwidth]{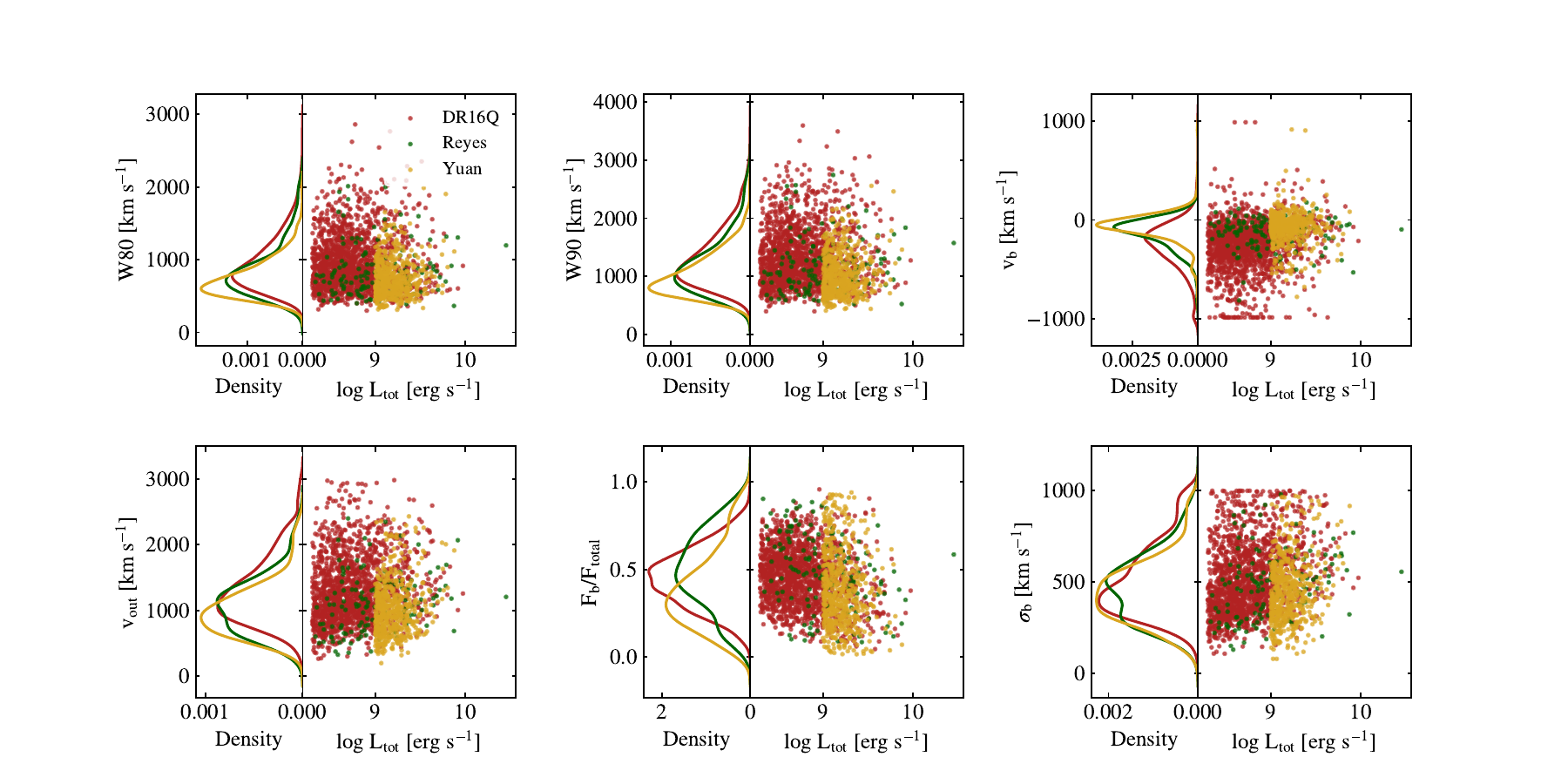}
    \caption{Relation between the measured kinematic properties and the luminosity of the QSOs 1 and 2. The QSOs 1 are shown as red dots, the QSOs 2 from the Reyes sample are shown in green and those from the Yuan sample are shown in yellow.}
    \label{fig:scatterplot_kin_Lum}
\end{figure*}

\bsp	
\label{lastpage}
\end{document}